\documentclass[10pt,twocolumn,letterpaper]{article}

\usepackage{cvpr}
\usepackage{times}
\usepackage{epsfig}
\usepackage{graphicx}
\usepackage{amsmath}
\usepackage{amssymb}

\usepackage{tikz} 
\usepackage{pgfplots}
\pgfplotsset{compat=1.5}

\usepackage{subfigure}
\usepackage{caption}
\usepackage{subfigure}
\usepackage{color}
\usepackage{placeins}
\usepackage{float}
\usepackage{tabularx,colortbl}
\usepackage{times}
\usepackage{epstopdf}
\usepackage{mathrsfs}
\usepackage{diagbox}
\usepackage{multirow}
\usepackage{algorithm}
\usepackage{algorithmic}
\usepackage{galois}
\usepackage{url}
\usepackage{amssymb}

\usepackage{relsize}

\usepackage[pagebackref=true,breaklinks=true,letterpaper=true,colorlinks,bookmarks=false]{hyperref}

\cvprfinalcopy 

\def\cvprPaperID{5734} 

\ifcvprfinal\fi
\begin{document}

\title{Learned Image Compression with Discretized Gaussian Mixture Likelihoods and Attention Modules}

\author{
Zhengxue Cheng$^{1}$, Heming Sun$^{2,3}$, Masaru Takeuchi$^{2}$, Jiro Katto$^{1}$ \\
\small{$^{1}$ Department of Computer Science and Communication Engineering, Waseda University, Tokyo, Japan}\\
\small{$^{2}$ Waseda Research Institute for Science and Engineering, Tokyo, Japan $^{3}$ JST, PRESTO, 4-1-8 Honcho, Kawaguchi, Saitama, Japan}\\
{\tt\small \{zxcheng@asagi., hemingsun@aoni., masaru-t@aoni., katto@\}waseda.jp}
}


\maketitle

\begin{abstract}
Image compression is a fundamental research field and many well-known compression standards have been developed for many decades. Recently, learned compression methods exhibit a fast development trend with promising results. However, there is still a performance gap between learned compression algorithms and reigning compression standards, especially in terms of widely used PSNR metric. In this paper, we explore the remaining redundancy of recent learned compression algorithms. We have found accurate entropy models for rate estimation largely affect the optimization of network parameters and thus affect the rate-distortion performance. Therefore, in this paper, we propose to use discretized Gaussian Mixture Likelihoods to parameterize the distributions of latent codes, which can achieve a more accurate and flexible entropy model. Besides, we take advantage of recent attention modules and incorporate them into network architecture to enhance the performance. Experimental results demonstrate our proposed method achieves a state-of-the-art performance compared to existing learned compression methods on both Kodak and high-resolution datasets. To our knowledge our approach is the first work to achieve comparable performance with latest compression standard Versatile Video Coding (VVC) regarding PSNR. More importantly, our approach generates more visually pleasant results when optimized by MS-SSIM. The project page is at \url{https://github.com/ZhengxueCheng/Learned-Image-Compression-with-GMM-and-Attention}.
\end{abstract}


\section{Introduction}

Image compression is an important and fundamental research topic in the field of signal processing for many decades to achieve efficient image transmission and storage. Classical image compression standards include JPEG~\cite{IEEEexample:JPEG}, JPEG2000~\cite{IEEEexample:JPEG2000}, HEVC/H.265~\cite{IEEEexample:HEVC} and ongoing Versatile Video Coding (VVC)~\cite{IEEEexample:VVC} which will be the next generation compression standard expected by the end of 2020. Typically they rely on hand-crafted creativity to present module-based encoder/decoder (codec) block diagrams. They use fixed transform matrix, intra prediction, quantization, context adaptive arithmetic coders and various de-blocking or loop filters to reduce spatial redundancy to improve the coding efficiency. The standardization of a conventional codec has historically spanned several years. Along with the fast development of new image formats and the proliferation of high-resolution mobile devices, existing image compression standards are not expected to be an optimal and general solutions for all kinds of image contents.

\begin{figure}[tb]
	\centerline{\psfig{figure=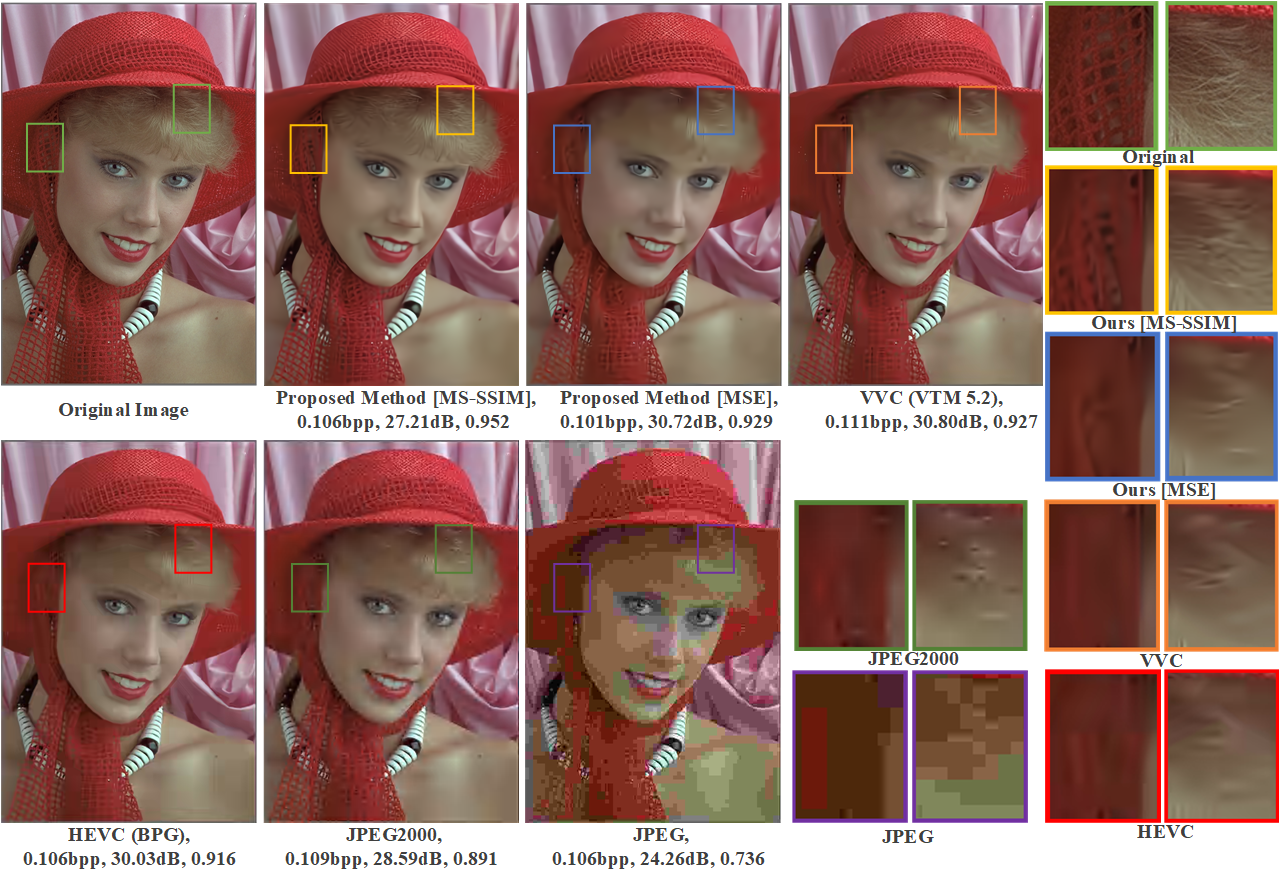,width=82 mm} }
	\caption{Visualization of reconstructed images \emph{Kodim04} from Kodak dataset with approximately $0.1$ bpp.}
	\label{fig:visualization04}
\end{figure}

Various approaches has been investigated for end-to-end image compression. Recently, the notable approaches are context-adaptive entropy models for learned image compression~\cite{IEEEexample:Balle2, IEEEexample:David, IEEEexample:Lee} to achieve superior performance among all the learned codecs. The work~\cite{IEEEexample:Balle2} proposed a hyperprior to add additional bits to model the entropy model. The work~\cite{IEEEexample:David} jointly combine an autoregressive mask convolution and the hyperprior. The work~\cite{IEEEexample:Lee} proposed a quite similar idea by considering two types of contexts, bit-consuming contexts (that is, hyperprior) and bit-free contexts (that is, autoregressive model) to realize a context-adaptive entropy model. Our method are based on the development of these recent entropy model techniques to further improve the performance.

In this paper, our main contribution is to present a more accurate and flexible entropy model by leveraging discretized Gaussian mixture likelihoods. We visualize the spatial redundancy of compressed codes from recent learned compression techniques. Motivated from it, we propose to use discretized Gaussian mixture likelihoods to parameterize the distributions, which removes remaining redundancy to achieve accurate entropy model, and thus directly lead to fewer required encoding bits. Besides, we adopt a simplified version of attention module into our network architecture. Attention module can make learned models pay more attention to complex regions to improve our coding performance with moderate training complexity.

Experimental results demonstrate our proposed method leads to the state-of-the-art performance on both PSNR and MS-SSIM quality metrics, in comparison with classical image compression standards HEVC, JPEG2000, JPEG and existing deep learning based compression approaches. To our knowledge, we are the first work to reach very close performance with ongoing versatile video coding test model VTM 5.2 with intra profile in terms of PSNR. Moreover, our method produces visually pleasant reconstructed images when optimizing by MS-SSIM as Fig.~\ref{fig:visualization04}.

\section{Related Work}

\noindent\textbf{Hand-crafted Compression} $\,$ Existing image compression standards, such as JPEG~\cite{IEEEexample:JPEG}, JPEG2000~\cite{IEEEexample:JPEG2000}, HEVC~\cite{IEEEexample:HEVC} and VVC~\cite{IEEEexample:VVC}, reply on hand-crafted module design individually. Specifically, these modules include intra prediction, discrete cosine transform or wavelet transform, quantization, and entropy coder such as Huffman coder or content adaptive binary arithmetic coder (CABAC). They design each module with multiple modes and conduct the rate-distortion optimization to determine the best mode. Especially, VVC~\cite{IEEEexample:VVC} supports larger coding unit, more prediction modes, more transform types and other coding tools. Besides, along with the development of classical compression algorithms, some hybrid methods have been proposed, by taking advantage of both conventional compression algorithms and latest learned super resolution approaches, such as~\cite{IEEEexample:CLIC}.

\noindent\textbf{Learned Compression} $\,$ Recently, we have seen a great surge of deep learning based image compression approaches utilizing autoencoder architecture~\cite{IEEEexample:autoencoder}, which have achieved a great success with promising results. The development of previous learning based compression models have spanned several years and have many related works. In the early stage, some works are proposed to deal with non-differential quantization and rate estimation to make end-to-end training possible, such as~\cite{IEEEexample:Theis, IEEEexample:Balle, IEEEexample:softQuan}. After that, some works focus on the design of network structure, which is able to extract more compact and efficient latent representation and to reconstruct high-quality images from compressed features. For instance, some works~\cite{IEEEexample:Toderici01, IEEEexample:Toderici, IEEEexample:Nick} use recurrent neural networks to compress the residual information recursively, but they mainly relied on binary representation at each iteration to achieve scalable coding for image compression. Some approaches~\cite{IEEEexample:waveone, IEEEexample:MITgan, IEEEexample:Extreme} use generative models to learn the distribution of images using adversarial training to achieve better subjective quality at extremely low bit rate. Some approaches include a content-weighted strategy~\cite{IEEEexample:HKPU} or de-correlating different channels using principle component analysis~\cite{ IEEEexample:PCS}, or consider energy compaction~\cite{IEEEexample:ourCVPR}, or use deep residual units to enhance the network architecture~\cite{IEEEexample:ourCLIC}. Recently, several studies investigate the adaptive context model for entropy estimation to navigate the optimization process of neural network parameters to achieve best tradeoff between reconstruction errors and required bits (entropy), including~\cite{IEEEexample:Balle2, IEEEexample:David, IEEEexample:Lee, IEEEexample:conditional}. Entropy estimation techniques have greatly improved the learned compression algorithms, and the most representative methods are hyperprior model and joint model. However, there is still a gap between the estimated distribution and true marginal distribution of latent representation.

\noindent\textbf{Parameterized Model}$\,$ Some image generation studies have investigated several parameterized distribution models. For instance, standard PixelCNN~\cite{IEEEexample:pixelcnn} uses full 256-way softmax likelihoods, but they are extremely memory-consuming. To address this problem, PixelCNN++~\cite{IEEEexample:pixelcnnplus} proposed a discretized logistic mixture likelihoods to achieve faster training. Learned lossless image compression work L3C~\cite{IEEEexample:l3c} followed the PixelCNN++ to use a logistic mixture model. These works are basically used to estimate the likelihoods of 8-bit pixel values with fixed range of $[0, 255]$. However, in learned image compression task, few studies explore the effect of parameterized distribution models.

\section{Proposed Method}

\subsection{Formulation of Learned Compression Models}

\begin{figure*}[tb]
\centering
\subfigure[Baseline]{
\label{Fig.op.1}
\includegraphics[width=0.18\textwidth]{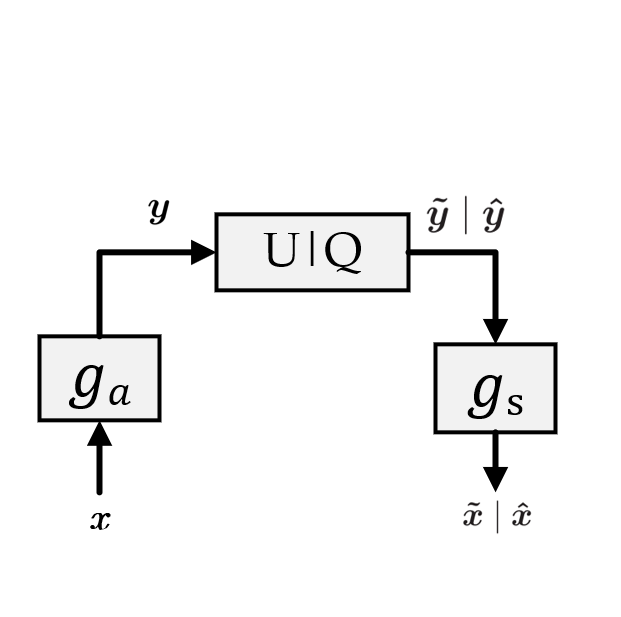}}
\hspace{0.05in}
\subfigure[Hyperprior]{
\label{Fig.op.2}
\includegraphics[width=0.22\textwidth]{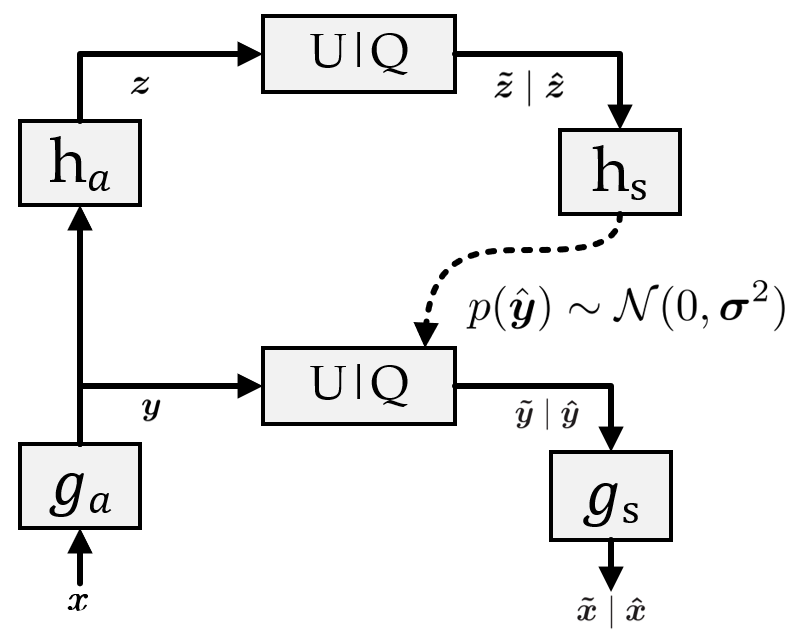}}
\hspace{0.05in}
\subfigure[Joint]{
\label{Fig.op.3}
\includegraphics[width=0.22\textwidth]{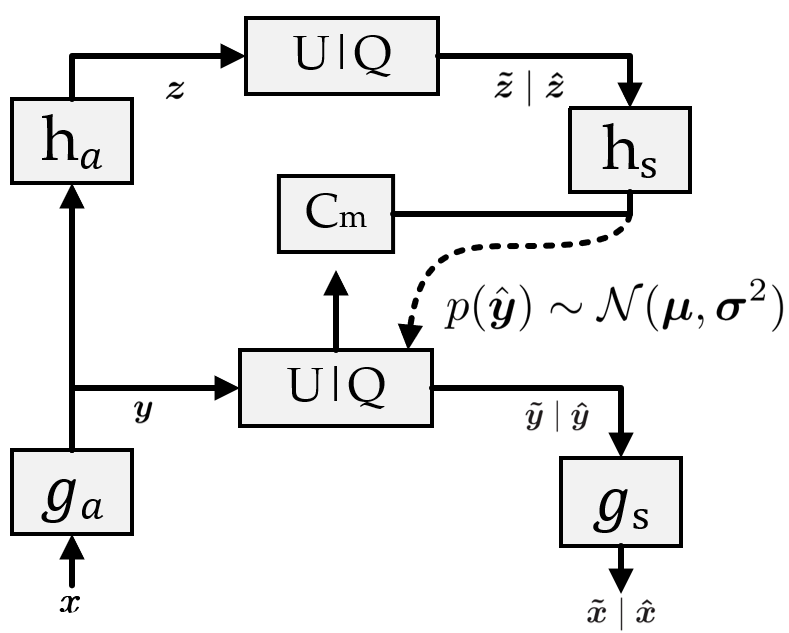}}
\subfigure[Proposed Model\;\;\;\;]{
\label{Fig.op.4}
\includegraphics[width=0.32\textwidth]{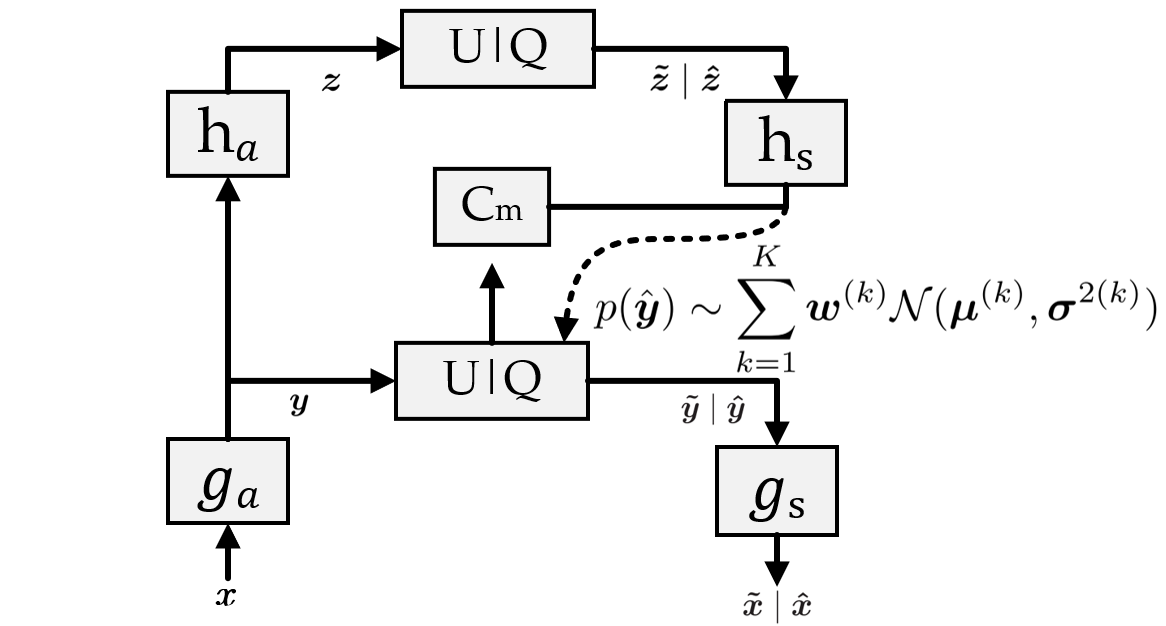}}
\caption{Operational diagrams of learned compression models (a)(b)(c) and proposed Gaussian Mixture Likelihoods (d).}
\label{fig:op}
\end{figure*}

In the transform coding approach~\cite{IEEEexample:transformcoding}, image compression can be formulated by (as Fig.~\ref{Fig.op.1})
\begin{equation}
\begin{aligned}
\boldsymbol{y} &=g_{a}(\boldsymbol{x}; \boldsymbol{\phi}) \\
\hat{\boldsymbol{y}} &= Q(\boldsymbol{y}) \\
\hat{\boldsymbol{x}} &=g_{s}(\hat{\boldsymbol{y}}; \boldsymbol{\theta})
\end{aligned}
\end{equation}
where $\boldsymbol{x}$, $\hat{\boldsymbol{x}}$, $\boldsymbol{y}$, and $\hat{\boldsymbol{y}}$ are raw images, reconstructed images, a latent presentation before quantization, and compressed codes, respectively. $\boldsymbol{\phi}$ and $\boldsymbol{\theta}$ are optimized parameters of analysis and synthesis transforms. $U|Q$ represents the quantization and entropy coding. During the training, the quantization is approximated by a uniform noise $\mathcal{U}(-\frac{1}{2}, \frac{1}{2})$ to generate noisy codes $\tilde{\boldsymbol{y}}$. During the inference, $U|Q$ represents real round-based quantization to generate $\hat{\boldsymbol{y}}$ and followed entropy coders to generate the bitstream. To simplify, we use $\hat{\boldsymbol{y}}$ to denote $\tilde{\boldsymbol{y}}|\hat{\boldsymbol{y}}$. If a probability model $p_{\hat{\boldsymbol{y}}}(\hat{\boldsymbol{y}})$ is given, entropy coding techniques, such as arithmetic coding~\cite{IEEEexample:arithmeticcoding}, can losslessly compress the quantized codes. Besides, arithmetic coder are a near-optimal entropy coder, which makes it feasible to use the entropy of $\boldsymbol{y}$ as the rate estimation during the training. In the baseline architecture, the marginal distribution of latent $\boldsymbol{y}$ is unknown and no additional bits are available to estimate $p_{\hat{\boldsymbol{y}}}(\hat{\boldsymbol{y}})$. Typically a non-adaptive density model is used and shared between encoder and decoder, also called factorized prior.

In the work~\cite{IEEEexample:Balle2}, \emph{Ball\'e} proposed a hyperprior, by introducing a side information $\boldsymbol{z}$ to capture spatial dependencies among the elements of $\boldsymbol{y}$, formulated by (as Fig.~\ref{Fig.op.2})
\begin{equation}
\begin{aligned}
\boldsymbol{z} &=h_{a}(\boldsymbol{y}; \boldsymbol{\phi_{h}}); \;\;\\
\hat{\boldsymbol{z}} &= Q(\boldsymbol{z}) \\
p_{\hat{\boldsymbol{y}}|\hat{\boldsymbol{z}}}(\hat{\boldsymbol{y}}|\hat{\boldsymbol{z}}) &\leftarrow h_{s}(\hat{\boldsymbol{z}}; \boldsymbol{\theta_{h}})\\
\end{aligned}
\end{equation}
where $h_{a}$ and $h_{s}$ denote the analysis and synthesis transforms in the auxiliary autoencoder, where $\boldsymbol{\phi_{h}}$ and $\boldsymbol{\theta_{h}}$ are optimized parameters. $p_{\hat{\boldsymbol{y}}|\hat{\boldsymbol{z}}}(\hat{\boldsymbol{y}}|\hat{\boldsymbol{z}})$ are estimated distributions conditioned on $\hat{\boldsymbol{z}}$. For instance, the work~\cite{IEEEexample:Balle2} parameterized a zero-mean Gaussian distribution with scale parameters $\boldsymbol{\sigma^{2}} = h_{s}(\hat{\boldsymbol{z}}; \boldsymbol{\theta_{h}})$ to estimate $p_{\hat{\boldsymbol{y}}|\hat{\boldsymbol{z}}}(\hat{\boldsymbol{y}}|\hat{\boldsymbol{z}})$.

Following that, an enhanced work~\cite{IEEEexample:David} proposed a more accurate entropy model, which jointly utilize an autoregressive context model (denoted as $C_{m}$ in Fig.~\ref{Fig.op.3}) and a mean and scale hyperprior. The work~\cite{IEEEexample:Lee} also proposed a similar idea. The operational diagram is explained in Fig.~\ref{Fig.op.3}.

Learned image compression is a Lagrangian multiplier-based rate-distortion optimization. The loss function is
\begin{equation}
\begin{aligned}
\mathcal{L} =& \mathcal{R(\hat{\boldsymbol{y}})} + \mathcal{R(\hat{\boldsymbol{z}})} + \lambda\cdot\mathcal{D(\boldsymbol{x}, \boldsymbol{\hat{x}})}\\
=& \mathop{\mathbb{E}}[-\log_{2}(p_{\hat{\boldsymbol{y}}|\hat{\boldsymbol{z}}}(\hat{\boldsymbol{y}}|\hat{\boldsymbol{z}}  ))] + \mathop{\mathbb{E}}[-\log_{2}(p_{\hat{\boldsymbol{z}}|\boldsymbol{\psi}}(\hat{\boldsymbol{z}}|\boldsymbol{\psi} ))]  \\
&+\lambda\cdot\mathcal{D(\boldsymbol{x}, \boldsymbol{\hat{x}})}
\end{aligned}
\end{equation}
where $\lambda$ controls the rate-distortion tradeoff. Different $\lambda$ values are corresponding to different bit rates. $\mathcal{D(\boldsymbol{x}, \boldsymbol{\hat{x}})}$ denotes the distortion term. There is no prior for $\hat{\boldsymbol{z}}$, so a factorized density model $\boldsymbol{\psi}$ is used to encode $\hat{\boldsymbol{z}}$ as
\begin{equation}
\begin{aligned}
&p_{\hat{\boldsymbol{z}}|\boldsymbol{\psi}}(\hat{\boldsymbol{z}}|\boldsymbol{\psi}) = \prod_{i} (p_{z_{i}|\boldsymbol{\psi}}(\boldsymbol{\psi})\ast \mathcal{U}(-\frac{1}{2}, \frac{1}{2}))(\hat{z}_{i}) \\
\end{aligned}
\end{equation}
where $z_{i}$ denotes the $i$-th element of $\boldsymbol{z}$, and $i$ specifies to the position of each element or each signal. The remaining part is how to model the $p_{\hat{\boldsymbol{y}}}(\hat{\boldsymbol{y}}|\hat{\boldsymbol{z}})$ accurately.

\begin{figure*}[tb]
	\centerline{\psfig{figure=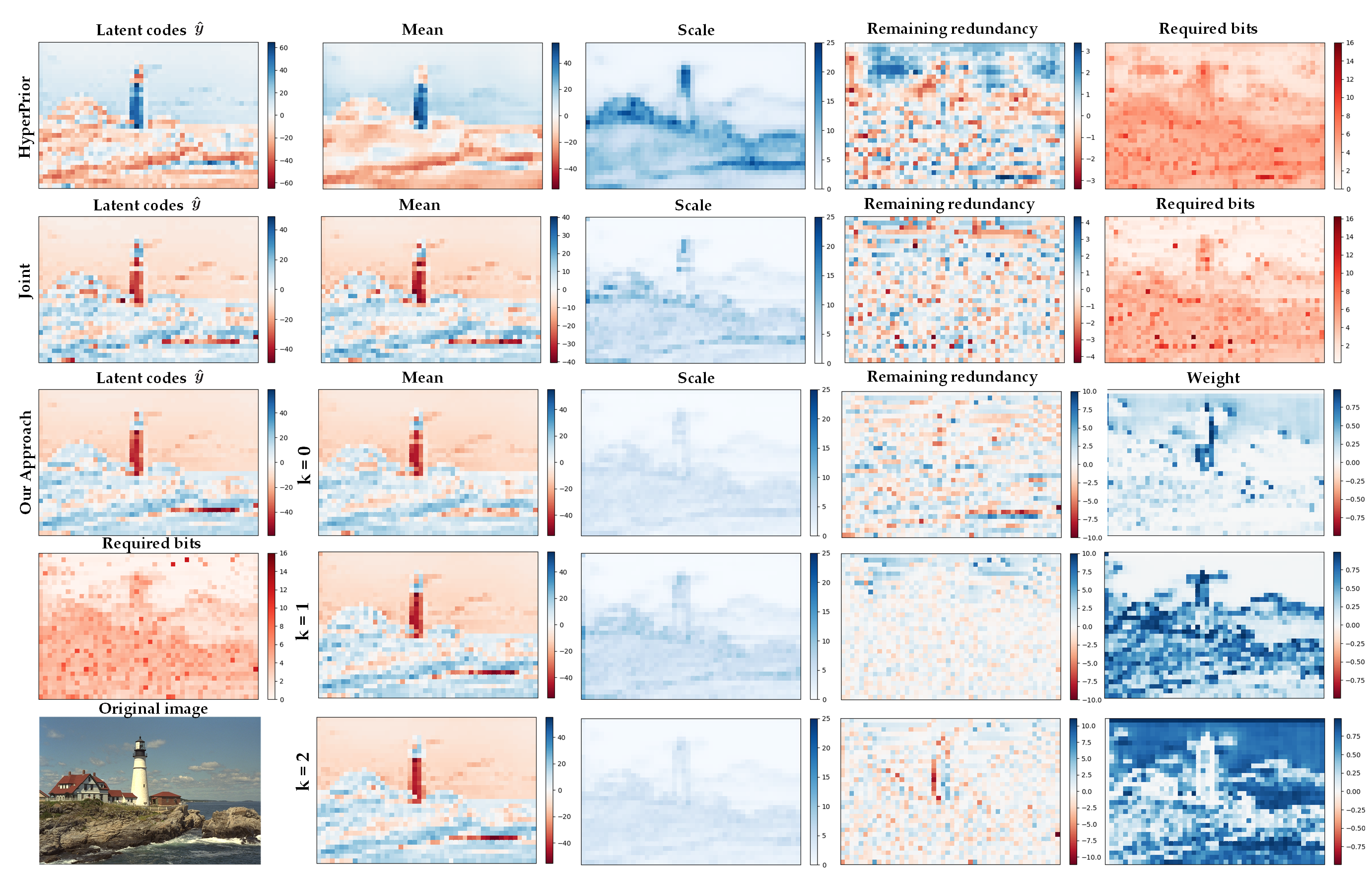, width=180 mm} }
	\caption{Visualization of different entropy models for the channel with the highest entropy using \emph{kodim21} from Kodak dataset as an example. It shows our approach provides a more flexible parameterized distribution models with smaller scale parameters and better spatial redundancy reduction, which directly results to a more accurate entropy model and fewer bits.}
	\label{fig:visualizationbits}
\end{figure*}

\subsection{Discretized Gaussian Mixture Likelihoods}


Whether the parameterized distribution model fits the marginal distribution of $\hat{\boldsymbol{y}}$ is a significant factor for the entropy model, thus affecting rate-distortion performance. Lots of efforts have been conducted to model the conditional probability distribution $p_{\hat{\boldsymbol{y}}}(\hat{\boldsymbol{y}}|\hat{\boldsymbol{z}})$ after decoding $\hat{\boldsymbol{z}}$. \emph{Ball\'e} work~\cite{IEEEexample:Balle2} firstly assumed a univariate Gaussian distribution model for the hyperprior, that is,
\begin{equation}
p_{\hat{\boldsymbol{y}}|\hat{\boldsymbol{z}}}(\hat{\boldsymbol{y}}|\hat{\boldsymbol{z}}) \sim \mathcal{N}(0, \boldsymbol{\sigma}^{2})
\end{equation}
The improved work~\cite{IEEEexample:David} extended a scale hyperprior to a mean and scale Gaussian distribution, that is,
\begin{equation}
p_{\hat{\boldsymbol{y}}|\hat{\boldsymbol{z}}}(\hat{\boldsymbol{y}}|\hat{\boldsymbol{z}}) \sim \mathcal{N}(\boldsymbol{\mu}, \boldsymbol{\sigma}^{2})
\end{equation}
and combine with a autoregressive model, denoted as \emph{Joint}.

To illustrate how entropy models work, we visualize different entropy models in Fig.~\ref{fig:visualizationbits} with the same network architecture. We only depict the channel with the highest entropy. The first rows visualizes the mean and scale~\emph{HyperPrior}. The $1$-st column is the quantized codes $\hat{\boldsymbol{y}}$. The second and third columns are the predicted mean $\boldsymbol{\mu}$ and scale $\boldsymbol{\sigma}$.
The $4$-th column is normalized values, calculate by $\frac{\hat{\boldsymbol{y}} - \boldsymbol{\mu} }{\boldsymbol{\sigma}}$. It is used to visualize the extent of remaining redundancy which is not captured by entropy models. The $5$-th column visualizes the \emph{Required bits} for each element for encoding, which is calculated as $(-\log_{2}(p_{\hat{\boldsymbol{y}}|\hat{\boldsymbol{z}}}(\hat{\boldsymbol{y}}|\hat{\boldsymbol{z}} )))$, which uses predicted distribution models.

Typically, predicted mean $\boldsymbol{\mu}$ is close to $\hat{\boldsymbol{y}}$. Complex regions have large $\boldsymbol{\sigma}$, requiring more bits for encoding. Instead, smooth regions have smaller $\boldsymbol{\sigma}$, resulting to fewer bits for encoding. \emph{HyperPrior} still has spatial redundancy remaining in simple regions, such as sky of \emph{kodim21}. Similarly, we visualize the \emph{Joint} entropy model. Compared with~\emph{HyperPrior}, \emph{Joint} removes more structures for normalized values by adding the autoregressive model, which is implemented by a $5\times5$ mask convolution, to capture the correlation with neighboring elements.

\begin{figure*}[tb]
	\centerline{\psfig{figure=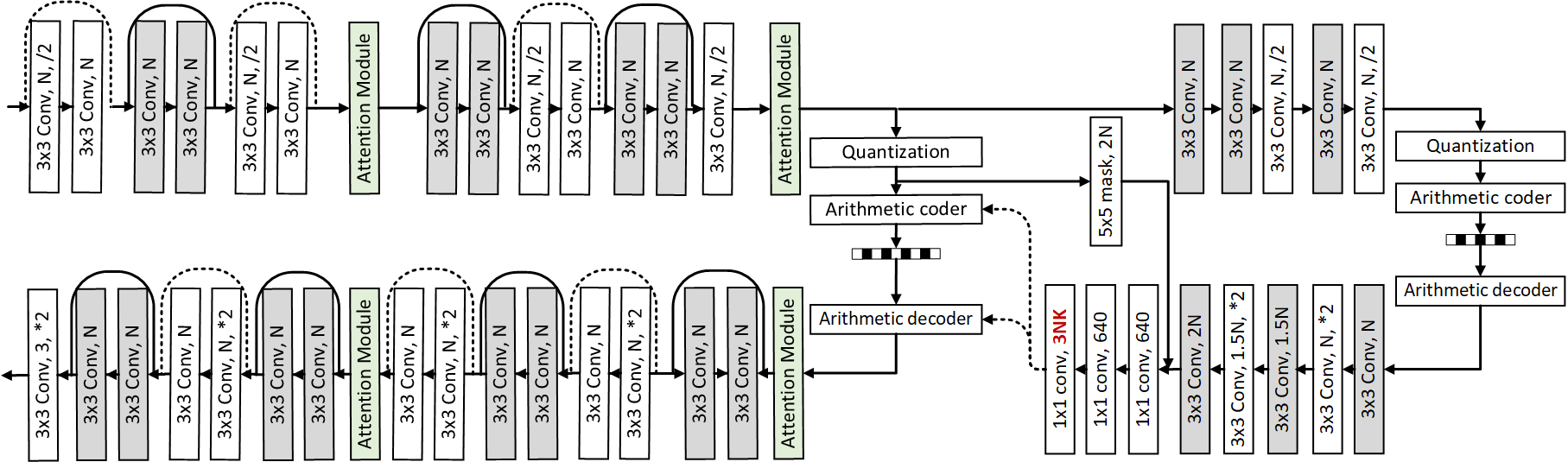,width=178mm} }
	\caption{Network architecture.}
	\label{fig:net}
\end{figure*}

However, \emph{Joint} entropy model is not perfect, because some spatial redundancy is still observed in the $2$-nd row, the $4$-th column of Fig.~\ref{fig:visualizationbits}. Although neighboring elements have already served as the input of context models, parameterized distributions cannot represent well to fully utilize the contexts and information from neighboring elements and additional bits $\boldsymbol{\hat{z}}$. It might be limited by fixed shape of single Gaussian distribution. This motivates us to consider more flexible parameterized models to achieve arbitrary likelihoods. Therefore, we propose the Gaussian mixture model, i.e.
\begin{equation}
\label{eq.continuous}
p_{\hat{\boldsymbol{y}}|\hat{\boldsymbol{z}}}(\hat{\boldsymbol{y}}|\hat{\boldsymbol{z}}) \sim \sum_{k=1}^{K}\boldsymbol{w}^{(k)}\mathcal{N}(\boldsymbol{\mu}^{(k)}, \boldsymbol{\sigma}^{2(k)})
\end{equation}
Eq.(\ref{eq.continuous}) usually refers to continuous values, but $\hat{\boldsymbol{y}}$ is discrete-valued after quantization. Inspiring by~\cite{IEEEexample:pixelcnnplus}, we propose to use discretized Gaussian mixture likelihoods. The reason why we did not use Logistic mixture likelihoods, because Gaussian achieves slightly better performance than logistic~\cite{IEEEexample:David}. Then the entropy model is formulated as
\begin{equation}
\begin{aligned}
p_{\hat{\boldsymbol{y}}|\hat{\boldsymbol{z}}}(\hat{\boldsymbol{y}}|\hat{\boldsymbol{z}}) &= \prod_{i} p_{\hat{\boldsymbol{y}}|\hat{\boldsymbol{z}}}(\hat{y}_{i}|\hat{\boldsymbol{z}})\\
p_{\hat{\boldsymbol{y}}|\hat{\boldsymbol{z}}}(\hat{y}_{i}|\hat{\boldsymbol{z}}) &= (\sum_{k=1}^{K}w_{i}^{(k)}\mathcal{N}(\mu_{i}^{(k)}, \sigma_{i}^{2(k)})\ast \mathcal{U}(-\frac{1}{2}, \frac{1}{2}))(\hat{y}_{i})\\
&=c(\hat{y}_{i} + \frac{1}{2}) - c(\hat{y}_{i} - \frac{1}{2})
\end{aligned}
\end{equation}
where $i$ specifies the location in feature maps. For example, $\hat{y}_{i}$ denotes the $i$-th element of $\boldsymbol{y}$ and $\mu_{i}$ denotes the $i$-th element of $\boldsymbol{\mu}$. $k$ denotes the index of mixtures. Each mixture is characterized by a Gaussian distribution with $3$ parameters, i.e. weights $w_{i}^{(k)}$, means $\mu_{i}^{(k)}$ and variances $\sigma_{i}^{2(k)}$ for each element $\hat{y}_{i}$. $c(\cdot)$ is the cumulative function. The range of $\boldsymbol{\hat{y}}$ is automatically learned and unknown ahead of time. To achieve stable training, we clip the range of $\hat{\boldsymbol{y}}$ to [-255, 256] because empirically $\hat{\boldsymbol{y}}$ would not exceed this range. For the edge case of $-255$, replace $c(\hat{y}_{i} - \frac{1}{2})$ by zero, i.e. $c(-\infty)=0$. For the edge case of $256$, replace $c(\hat{y}_{i} + \frac{1}{2})$ by one, i.e. $c(+\infty)=1$. It provides a numerically stable implementation for training.

The visualization of our approach is shown as the last three rows in Fig.~\ref{fig:visualizationbits}. We use $K=3$ in our experiments. Different from the above two rows, the $5$-th column shows the weights $w_{i}^{(k)}$ for each element and each mixture. Although each mixture still remains some spatial redundancy as shown in the $4$-th column in some parts, mixture model can adjust weights to different mixtures and different regions. For instance, the mixture $k=1$ has some redundancy in the sky as shown in the $4$-th row and $4$-th column, but the weights of this mixture for the sky are very small. The mixture $k=2$ remains some redundancy in the while tower as shown in the $5$-th row and $4$-th column, and mixture model also assigns small weights to the tower regions as shown in the $5$-th row and $5$-th column. Besides, the scales of our approach are smaller than~\emph{Joint} and~\emph{Hyperprior}, which demonstrates our entropy model is more accurate, thus directly resulting to fewer bits. \emph{Required bits} at the $5$-th row and $1$-st column visualizes the required bits of our approach for encoding, which is calculated by $(-\log_{2}(p_{\hat{\boldsymbol{y}}|\hat{\boldsymbol{z}}}(\hat{y}_{i}|\hat{\boldsymbol{z}}) ))$. Required bits of our approach is fewer than~\emph{Joint} as Table~\ref{table.bits} and visualized in Fig.~\ref{fig:visualizationbits}. All the models are trained with $\lambda=0.015$ with the same network architecture.

\begin{table}[tb]
\centering
\small
\caption{Required bits and quality metrics for \emph{Kodim21}.}
\label{table.bits}
\begin{tabular}{cccc}
\hline
Model & PSNR (dB) & MS-SSIM & Rate (bpp)  \\
\hline
\emph{Joint}  &33.435   &0.980    &0.533\\
\emph{Ours}  &33.623  &0.981    &0.519\\
\hline
\end{tabular}
\end{table}

The Gaussian mixture model can achieve better performance because it selects $K$ most probable values and assign small scale (i.e. high likelihoods) to them. Somehow this mechanism resembles the most probable symbol (MPS) in the Context-based Adaptive Binary Arithmetic coding (CABAC), which has been widely used in traditional video coding standards, but our latent codes are not limited to be binary. To the one extreme, if three selected mean values are the same, it will degrades to a single Gaussian model, which happens in the smooth regions. To the other extreme, three selected mean values are completely different from each other, our model would have three peaks of likelihoods representing three most probable values, which usually happens in the boundaries, edges or complex regions. Besides, all the parameters, weights $w_{i}^{(k)}$, means $\mu_{i}^{(k)}$ and variances $\sigma_{i}^{2(k)}$ for each element, are learnable during the training. Therefore, our method is a flexible and accurate entropy model, and operation diagram is shown as Fig.~\ref{Fig.op.4}.

\subsection{Network Architecture}

Our network architecture has a similar structure as~\cite{IEEEexample:ourCLIC} in Fig.~\ref{fig:net}. We use residual blocks to increase large receptive field and improve the rate-distortion performance. Decoder side uses subpixel convolution instead of transposed convolution as upsampling units to keep more details. $N$ denotes the number of channels and represents the model capacity. We use Gaussian mixture model, thus requiring $3\times N\times K$ channels for the output of auxiliary autoencoder.

\begin{figure}[tb]
\centering
\subfigure[Attention module]{
\label{Fig.attention.1}
\includegraphics[height = 1.3cm]{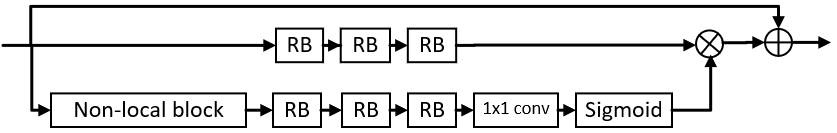}}
\subfigure[Simplified attention module]{
\label{Fig.attention.2}
\includegraphics[height = 1.3cm]{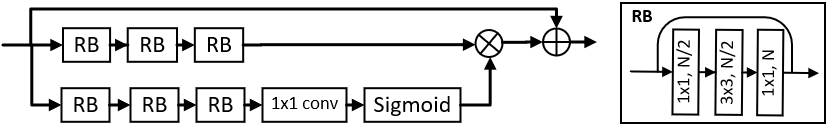}}
\caption{Different attention modules.}
\label{fig:attention}
\end{figure}

\begin{table}[tb]
\centering
\small
\caption{Performance of different attention modules.}
\label{table.attention}
\begin{tabular}{ccccc}
\hline
\emph{Module} & \emph{(a)w/ NLB}  & \emph{(b)w/o NLB}  & \emph{w/o attention}  \\
\hline
Loss                    &2.705    &2.754  &3.026\\
Time(s)/epoch  &1119     &336    &216\\
\hline
\end{tabular}
\end{table}

Furthermore, recent works use attention module to improve the performance for image restoration~\cite{IEEEexample:attention} and compression~\cite{IEEEexample:nonlocal}. The proposed attention module is illustrated in Fig.~\ref{Fig.attention.1}, but very time-consuming for training. We simplify this module by removing the non local block, because the deep residual blocks can already capture very large receptive field in our network architecture. The loss and training time comparison is given by Table~\ref{table.attention}, where this loss is trained after 16 epoches. Simplified attention module is shown in Fig.~\ref{Fig.attention.2} and can also reduce the loss with moderate complexity. Attention module can help the networks to pay more attention to challenging parts and reduce the bits of simple parts. Then we insert simplified attention module into encoder-decoder network as Fig.~\ref{fig:net}.

\section{Implementation Details}

\noindent\textbf{Training Details}$\quad$ For training, we used a subset of ImageNet database~\cite{IEEEexample:ImageNet}, and cropped them into 13830 samples with the size of $256\times 256$. To train our image compression models, the model was optimized using Adam~\cite{IEEEexample:adam} with a batch size of 8. The learning rate was maintained at a fixed value of $1\times10^{-4}$ during the training process, and was reduced to $1\times10^{-5}$ for the last $80k$ iterations.

We optimized our models using two quality metrics, i.e. mean square error (MSE) and MS-SSIM. When optimized by MSE, $\lambda$ belongs to the set $\{0.0016, 0.0032, 0.0075, 0.015, 0.03, 0.045\}$. $N$ is set as $128$ for three lower-rate models, and is set as $192$ for three higher rate models. To achieve high subjective quality, we also train the model using MS-SSIM quality metrics~\cite{IEEEexample:msssim} and then distortion term is defined by $\mathcal{D}(\boldsymbol{x}, \hat{\boldsymbol{x}}) = 1 - \rm{MS\text{-}SSIM}(\boldsymbol{x}, \hat{\boldsymbol{x}})$. When optimized by MS-SSIM, $\lambda$ is in the set $\{3,12,40,120\}$. $N$ is set as $128$ for two lower-rate models, and is set as $192$ for the two higher-rate models. Each model was trained up to $10^{6}$ iterations for each $\lambda$ to achieve stable performance.

\noindent\textbf{Evaluation}$\quad$ For comparison, we tested commonly used Kodak lossless image database~\cite{IEEEexample:kodak} with 24 uncompressed $768\times 512$ images. To validate the robustness of our proposed method, we also tested our proposed method using CVPR workshop CLIC professional validation dataset~\cite{IEEEexample:CLICdata} with $41$ high resolution and high quality images.

To evaluate the rate-distortion performance, the rate is measured by bits per pixel (bpp), and the quality is measured by either PSNR or MS-SSIM, corresponding to optimized distortion metrics. The rate-distortion (RD) curves are drawn to demonstrate their coding efficiency.

\noindent\textbf{Traditional Codecs}$\quad$ For VVC and HEVC, we used the official test model VTM 5.2 (accessed on July 2019)~\cite{IEEEexample:VTM} with intra profile and BPG software~\cite{IEEEexample:BPG} to test the performance. For both of them, we used non-default \emph{YUV444} format as the configuration, because it prevents the color component loss during color space conversion. We used official test model OpenJPEG~\cite{IEEEexample:Openjpeg} with default configuration $YUV420$ to represent the performance of JPEG2000. For JPEG compression, we use PIL library~\cite{IEEEexample:PIL}.

\vspace{-2mm}
\section{Experiments}


\begin{figure}[tb]
\centering
\subfigure[Loss curves with N=128.]{
\label{Fig.curve.1}
\includegraphics[height=3.8cm]{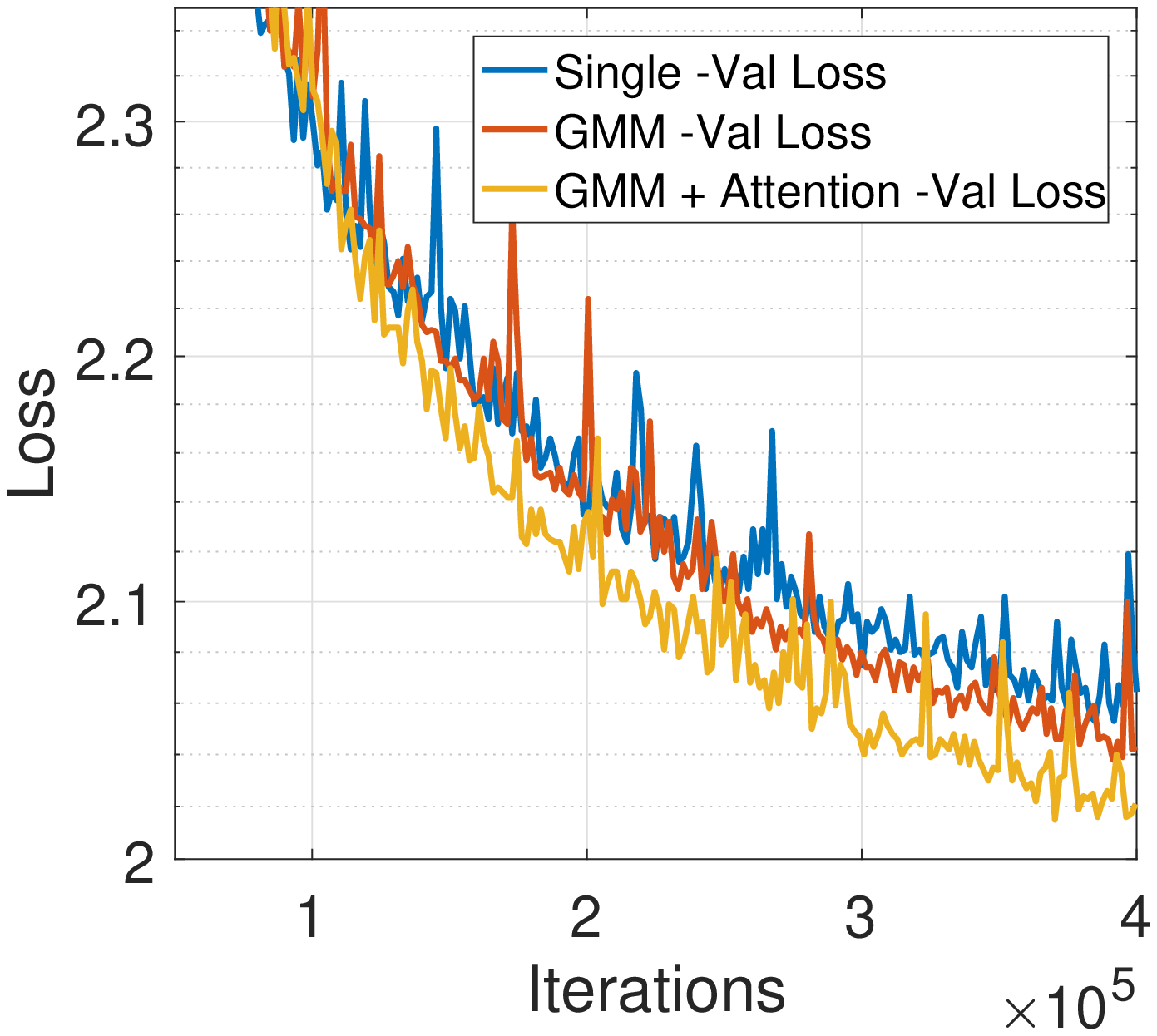}}
\subfigure[RD points with N=128.]{
\label{Fig.ablation.1}
\includegraphics[height=3.5cm]{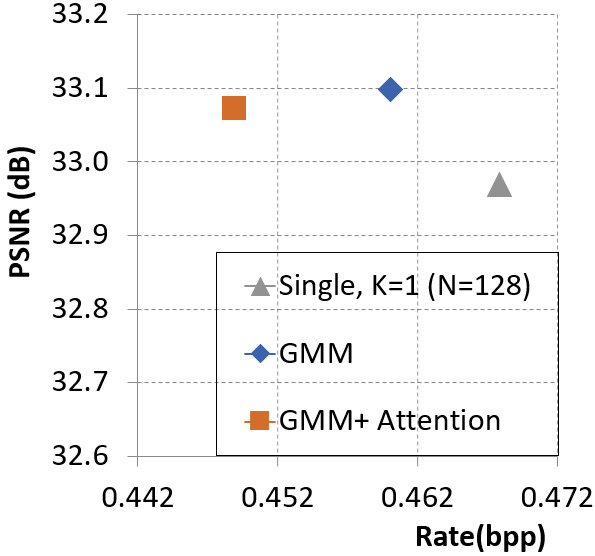}}
\subfigure[Loss curves with N=192.]{
\label{Fig.curve.2}
\includegraphics[height=3.8cm]{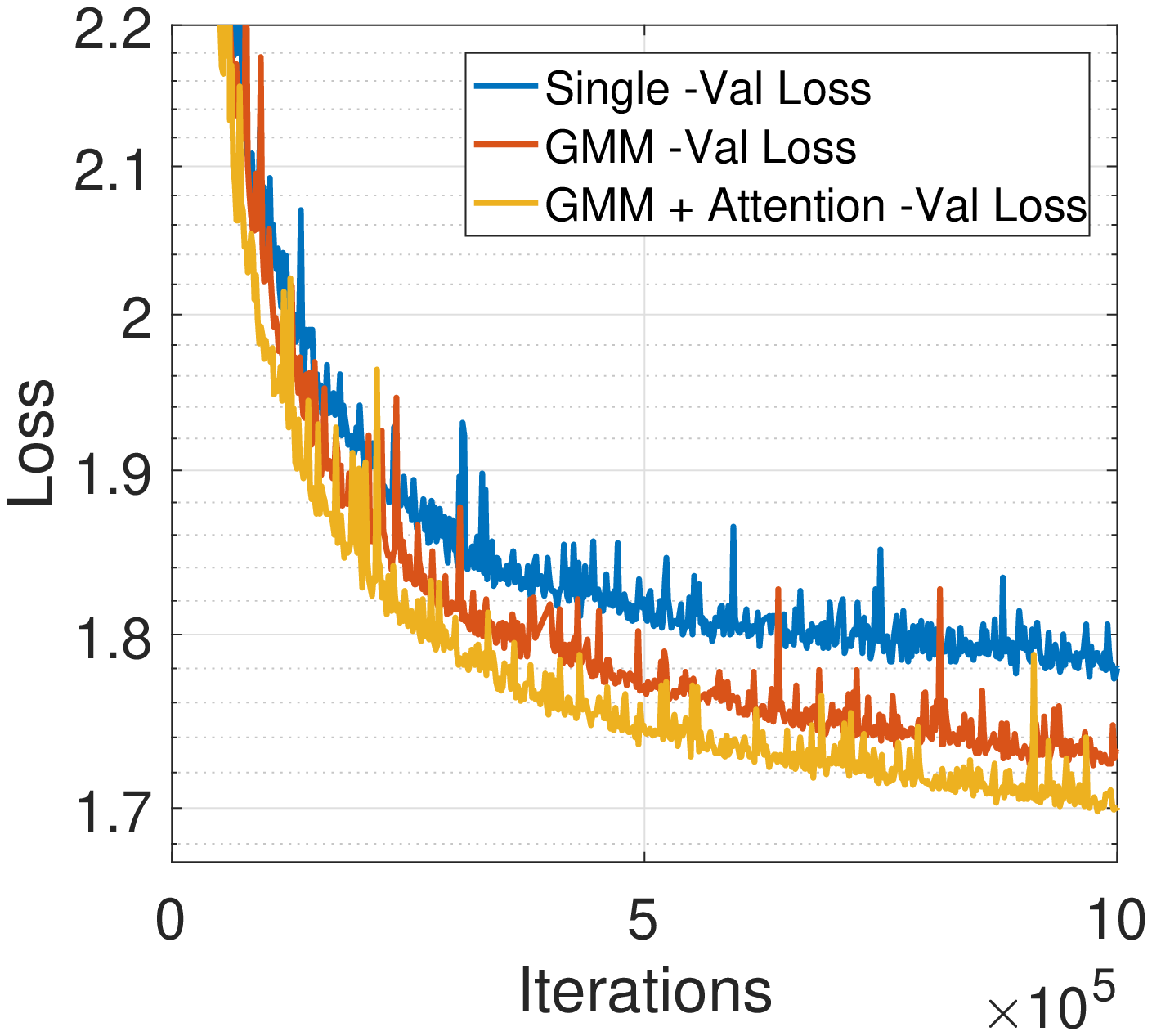}}
\subfigure[RD points with N=192.]{
\label{Fig.ablation.2}
\includegraphics[height=3.6cm]{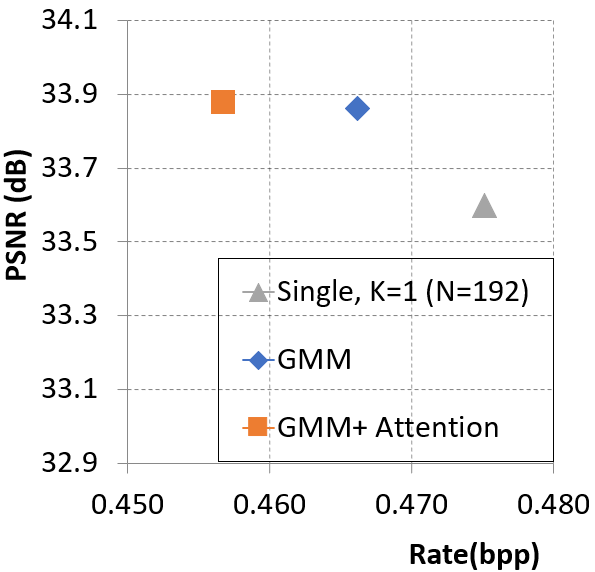}}
\caption{Ablation Study.}
\label{fig:losscurve}
\end{figure}

\begin{figure*}[tb]
\centering
\subfigure{
\pgfplotsset{
tick label style={font=\footnotesize},
label style={font=\footnotesize},
legend style={font=\footnotesize}
}
\pgfplotsset{every axis plot/.append style={line width=0.7pt}}
\tikzset{every mark/.append style={scale=0.5}}
\begin{tikzpicture}
\begin{axis}[
  width=8.5cm, height=7.0cm,
  xlabel = {Rate (bpp)},
  ylabel = {PSNR (dB)},
  xmin = 0.07,
  xmax = 0.9,
  ymin = 23.5,
  ymax = 37.5,
  minor y tick num = 1,
  grid = both,
  grid style = {gray!30},
  legend entries = {Proposed Method [MSE], Proposed Method [MS-SSIM], VVC-intra (VTM 5.2), Minnen [MSE] (NIPS18)~\cite{IEEEexample:David}, Lee [MSE] (ICLR19)~\cite{IEEEexample:Lee}, Ball\'e [MSE] (ICLR18)~\cite{IEEEexample:Balle2}, Ball\'e [MS-SSIM] (ICLR18)~\cite{IEEEexample:Balle2}, Li (CVPR2018)~\cite{IEEEexample:HKPU}, HEVC-intra (BPG), JPEG2000 (OpenJPEG), JPEG},
  legend style={font=\fontsize{5}{5}\selectfont, row sep=-3pt},
  legend cell align=left,
  legend pos = {south east},
  ]
  \addplot[smooth, blue!70!white, mark=*] table {data_Kodak_Proposed_MSE_2.dat};
  \addplot[smooth, blue!70!white, mark=square*] table {data_Kodak_Proposed_MSSSIM_2.dat};
  \addplot[smooth, orange] table {data_Kodak_VTM444_2.dat};
  \addplot[smooth, violet] table {data_Kodak_Minnen_2.dat};
  \addplot[smooth, gray] table {data_Kodak_Lee_2.dat};
  \addplot[smooth, purple] table {data_Kodak_Balle_MSE_2.dat};
  \addplot[smooth, yellow!70!red] table {data_Kodak_Balle_MSSSIM_2.dat};
  \addplot[smooth, olive] table {data_Kodak_HKPU_2.dat};
  \addplot[smooth, red!90!white] table {data_Kodak_BPG_2.dat};
  \addplot[smooth, green!60!black] table {data_Kodak_JPEG2K_2.dat};
  \addplot[smooth, magenta] table {data_Kodak_JPEG_2.dat};
\end{axis}
\end{tikzpicture}
}
\subfigure{
\pgfplotsset{
tick label style={font=\footnotesize},
label style={font=\footnotesize},
legend style={font=\footnotesize}
}
\pgfplotsset{every axis plot/.append style={line width=0.7pt}}
\tikzset{every mark/.append style={scale=0.5}}
\begin{tikzpicture}
\begin{axis}[
  width=8.5cm, height=7.0cm,
  xlabel = {Rate (bpp)},
  ylabel = {MS-SSIM (dB)},
  xmin = 0.07,
  xmax = 0.9,
  ymin = 8,
  ymax = 23,
  minor y tick num = 1,
  grid = both,
  grid style = {gray!30},
  legend entries = {Proposed Method [MSE], Proposed Method [MS-SSIM], VVC-intra (VTM5.2), Minnen [MS-SSIM] (NIPS18)~\cite{IEEEexample:David}, Lee [MS-SSIM] (ICLR19)~\cite{IEEEexample:Lee}, Ball\'e [MSE] (ICLR18)~\cite{IEEEexample:Balle2}, Ball\'e [MS-SSIM] (ICLR18)~\cite{IEEEexample:Balle2}, HEVC-intra (BPG), JPEG2000 (OpenJPEG), JPEG},
  legend style={font=\fontsize{5}{5}\selectfont, row sep=-3pt},
  legend cell align=left,
  legend pos = {south east},
  ]
  \addplot[smooth, blue!70!white, mark=*] table {data_Kodak_Proposed_MSE.dat};
  \addplot[smooth, blue!70!white, mark=square*] table {data_Kodak_Proposed_MSSSIM.dat};
  \addplot[smooth, orange] table {data_Kodak_VTM444.dat};
  \addplot[smooth, violet] table {data_Kodak_Minnen.dat};
  \addplot[smooth, gray] table {data_Kodak_Lee.dat};
  \addplot[smooth, purple] table {data_Kodak_Balle_MSE.dat};
  \addplot[smooth, yellow!70!red] table {data_Kodak_Balle_MSSSIM.dat};
  \addplot[smooth, red!90!white] table {data_Kodak_BPG.dat};
  \addplot[smooth, green!60!black] table {data_Kodak_JPEG2K.dat};
  \addplot[smooth, magenta] table {data_Kodak_JPEG.dat};
\end{axis}
\end{tikzpicture}
}
\vspace{-0.5cm}
\caption{Performance Evaluation on Kodak dataset.}
\vspace{-0.3cm}
\label{fig:kodak}
\end{figure*}
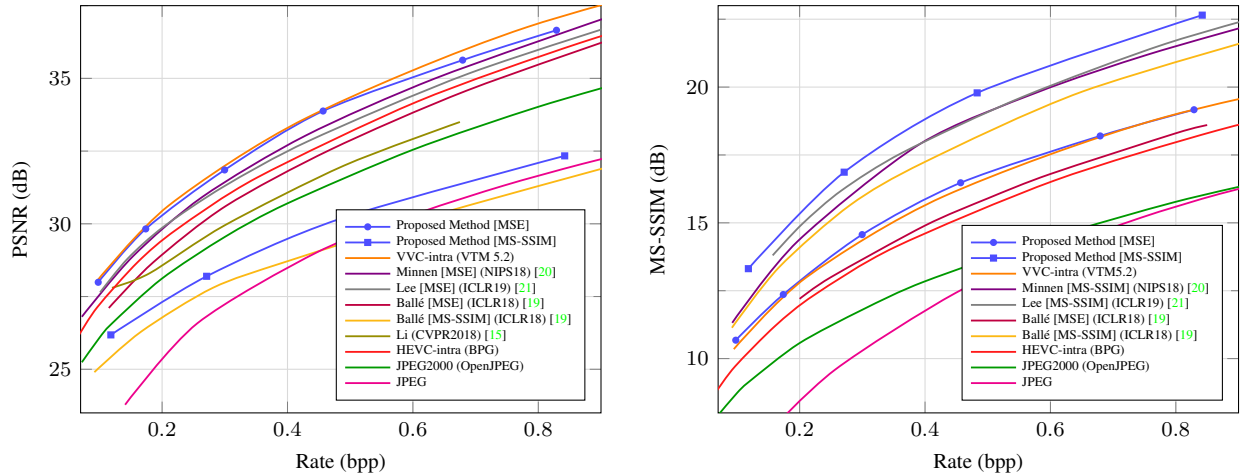

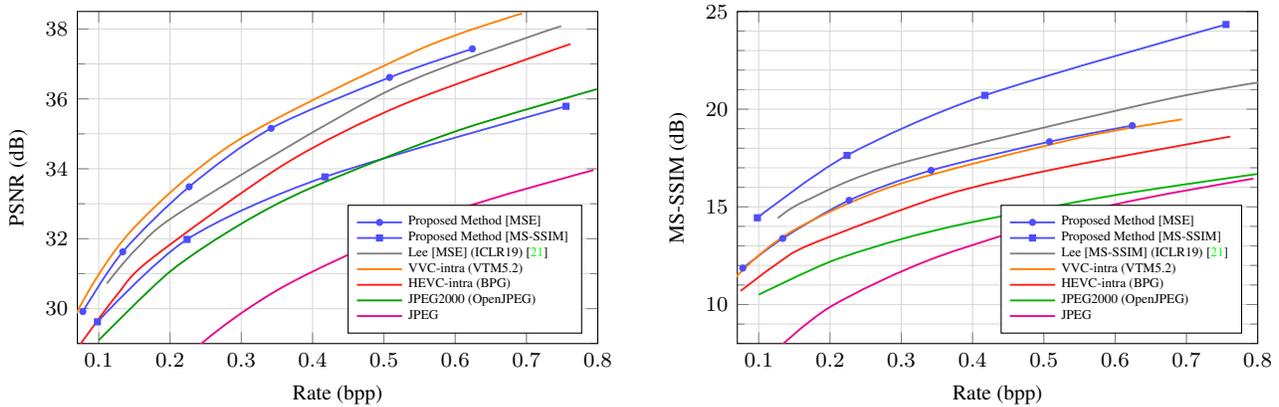
\begin{figure*}[tb]
\centering
\subfigure{
\pgfplotsset{
tick label style={font=\footnotesize},
label style={font=\footnotesize},
legend style={font=\footnotesize}
}
\pgfplotsset{every axis plot/.append style={line width=0.7pt}}
\tikzset{every mark/.append style={scale=0.5}}
\begin{tikzpicture}
\begin{axis}[
  width=8.5cm, height=6.0cm,
  xlabel = {Rate (bpp)},
  ylabel = {PSNR (dB)},
  xmin = 0.07,
  xmax = 0.8,
  ymin = 29,
  ymax = 38.5,
  minor y tick num = 1,
  grid = both,
  grid style = {gray!30},
  legend entries = {Proposed Method [MSE], Proposed Method [MS-SSIM], Lee [MSE] (ICLR19)~\cite{IEEEexample:Lee}, VVC-intra (VTM5.2), HEVC-intra (BPG), JPEG2000 (OpenJPEG), JPEG},
  legend style={font=\fontsize{5}{5}\selectfont, row sep=-3pt},
  legend cell align=left,
  legend pos = {south east},
  ]
  \addplot[smooth, blue!70!white, mark=*] table {data_CLIC_Proposed_MSE_2.dat};
  \addplot[smooth, blue!70!white, mark=square*] table {data_CLIC_Proposed_MSSSIM_2.dat};
  \addplot[smooth, gray] table {data_CLIC_Lee_2.dat};
  \addplot[smooth, orange] table {data_CLIC_VTM444_2.dat};
  \addplot[smooth, red!90!white] table {data_CLIC_BPG_2.dat};
  \addplot[smooth, green!60!black] table {data_CLIC_JPEG2K_2.dat};
  \addplot[smooth, magenta] table {data_CLIC_JPEG_2.dat};
\end{axis}
\end{tikzpicture}
}
\subfigure{
\pgfplotsset{
tick label style={font=\footnotesize},
label style={font=\footnotesize},
legend style={font=\footnotesize}
}
\pgfplotsset{every axis plot/.append style={line width=0.7pt}}
\tikzset{every mark/.append style={scale=0.5}}
\begin{tikzpicture}
\begin{axis}[
  width=8.5cm, height=6.0cm,
  xlabel = {Rate (bpp)},
  ylabel = {MS-SSIM (dB)},
  xmin = 0.07,
  xmax = 0.8,
  ymin = 8,
  ymax = 25,
  minor y tick num = 4,
  grid = both,
  grid style = {gray!30},
  legend entries = {Proposed Method [MSE], Proposed Method [MS-SSIM], Lee [MS-SSIM] (ICLR19)~\cite{IEEEexample:Lee}, VVC-intra (VTM5.2), HEVC-intra (BPG), JPEG2000 (OpenJPEG), JPEG},
  legend style={font=\fontsize{5}{5}\selectfont, row sep=-3pt},
  legend cell align=left,
  legend pos = {south east},
  ]
  \addplot[smooth, blue!70!white, mark=*] table {data_CLIC_Proposed_MSE.dat};
  \addplot[smooth, blue!70!white, mark=square*] table {data_CLIC_Proposed_MSSSIM.dat};
  \addplot[smooth, gray] table {data_CLIC_Lee.dat};
  \addplot[smooth, orange] table {data_CLIC_VTM444.dat};
  \addplot[smooth, red!90!white] table {data_CLIC_BPG.dat};
  \addplot[smooth, green!70!black] table {data_CLIC_JPEG2K.dat};
  \addplot[smooth, magenta] table {data_CLIC_JPEG.dat};
\end{axis}
\end{tikzpicture}
}
\vspace{-0.7cm}
\caption{Performance Evaluation on CLIC Professional Validation dataset.}
\label{fig:clic}
\end{figure*}

\subsection{Ablation Study}

In order to show the effectiveness of our proposed Gaussian mixture model (GMM) and simplified attention module, we test them with different model capacity, i.e. $N$=128, and $N$=192. These models are optimized by MSE with $\lambda=0.015$. The loss curves are shown in Fig.~\ref{Fig.curve.1} and Fig.~\ref{Fig.curve.2}. The asymptotic loss curves shows along with the increase of training iterations, the effectiveness of our proposed approaches become strong. The corresponding rate-distortion points are depicted in Fig.~\ref{Fig.ablation.1} and Fig.~\ref{Fig.ablation.2}, to show the coding gain of proposed approaches. It can be observed our proposed approaches can improve the rate-distortion performance regardless of the model capacity. Besides, GMM works better on $192$ filters than $128$ filters, probably because $192$ filters have large model capacity, so easily resulting more remaining spatial redundancy, and our proposed Gaussian mixture likelihoods can capture them effectively to reduce the bits.

\subsection{Rate-distortion Performance}

The rate-distortion performance on Kodak dataset is shown in Fig.~\ref{fig:kodak}. MS-SSIM is converted to decibels ($-10\log_{10}(1-$MS-SSIM$)$) to illustrate the difference clearly. We compare our method with well-known compression standards, and recent neural network-based learned compression methods, including the works of Li~\emph{et al.}~\cite{IEEEexample:HKPU}, Ball\'e~\emph{et al.}~\cite{IEEEexample:Balle2}, Minnen~\emph{et al.}~\cite{IEEEexample:David} and Lee~\emph{et al.}~\cite{IEEEexample:Lee}. RD curves of~\cite{IEEEexample:Lee} are from their released source code\footnote{\footnotesize{\url{https://github.com/JooyoungLeeETRI/CA_Entropy_Model}}}. RD points of \cite{IEEEexample:Balle2} and~\cite{IEEEexample:David} are obtained by contacting corresponding authors. RD points of~\cite{IEEEexample:HKPU} are traced from their paper with only PSNR. Regarding PSNR, our method yields a competitive results with VVC and achieves better coding performance than previous learning based methods. To our knowledge, our approach is the first work to achieve comparable PSNR with VVC. Regarding MS-SSIM, our method achieves state-of-the-art results among existing works.

Fig.~\ref{fig:clic} shows the comparison results with JPEG, JPEG2000, HEVC, VVC and the work of Lee \emph{et al.}~\cite{IEEEexample:Lee} on CLIC validation dataset. Regarding PSNR, our approach outperforms all the other codecs, except for VVC. This performance gap might be resulted from small size of training patches. Regarding MS-SSIM, our approaches significantly outperform all the codecs. It shows our method also works for high resolution images.

\begin{figure*}[tb]
	\centerline{\psfig{figure=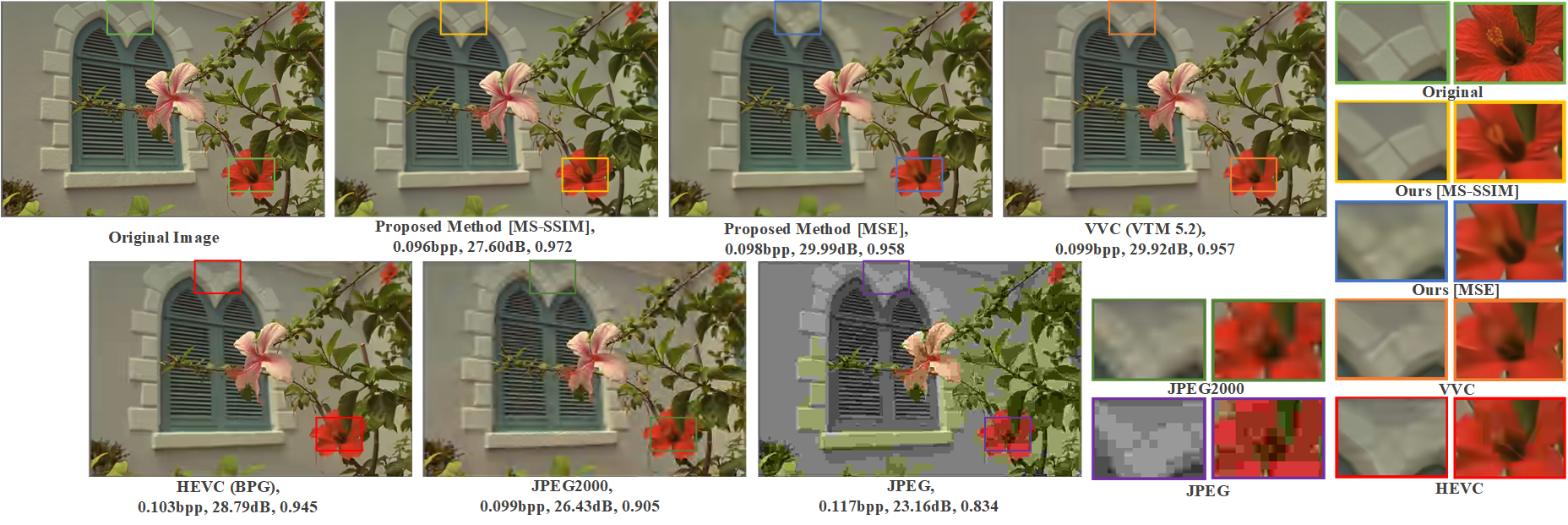,width=173 mm} }
	\caption{Visualization of reconstructed images \emph{kodim07} from Kodak dataset.}
	\label{fig:visualization07}
\end{figure*}

\begin{figure*}[tb]
	\centerline{\psfig{figure=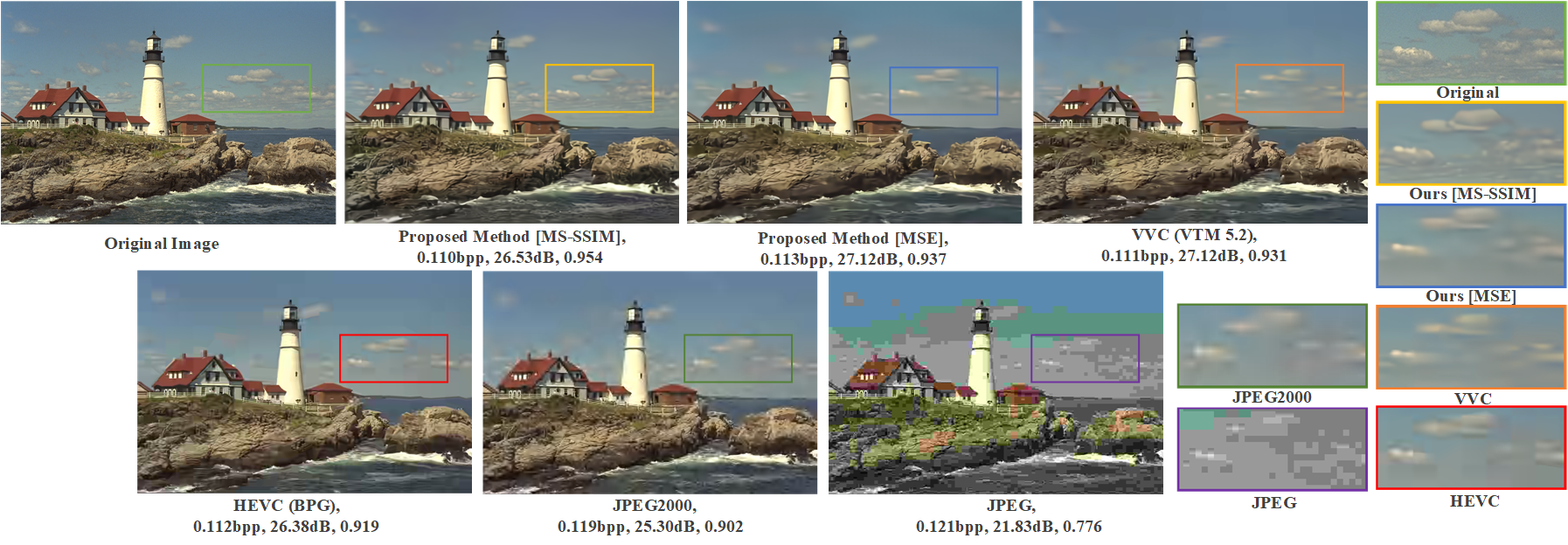,width=173 mm} }
	\caption{Visualization of reconstructed images \emph{kodim21} from Kodak dataset.}
	\label{fig:visualization21}
\end{figure*}

\subsection{Qualitative Results}

To demonstrate our method can generate more visually pleasant results, we visualize some reconstructed images for qualitative performance comparison.

Fig.~\ref{fig:visualization04} shows reconstructed images \emph{kodim04} with approximately $0.10$ bpp and a compression ratio of 240:1. More details are kept for our method optimized by MS-SSIM, and the hair appears much more natural than other codecs. Our method optimized by MSE achieves comparable results with VVC, and outperform HEVC, JPEG2000, JPEG. Fig.~\ref{fig:visualization07} shows reconstructed images \emph{kodim07} with approximately $0.10$ bpp and a compression ratio of 240:1. Our proposed method optimized by MS-SSIM generates more visually pleasant results, as shown by enlarged brick wall and red flowers. Our model optimized by MSE is slightly worse than VVC, but better than HEVC, JPEG2000 and JPEG. Fig.~\ref{fig:visualization21} shows reconstructed images \emph{kodim21} with approximately $0.12$ bpp and a compression ratio of 200:1. The shape of cloud is well preserved in our proposed method optimized by MS-SSIM. Our proposed method optimized by MSE achieves comparable quality with VVC. For the other three codecs, i.e. HEVC, JPEG2000 and JPEG, clear artifacts and blocking effect appeared. More subjective quality results are visualized in supplementary materials.

\section{Conclusion}

We propose a learned image compression approach using a discretized Gaussian mixture likelihoods and attention modules. By exploring the remaining redundancy of recent learned compression techniques, we have found single parameterized model can not achieve arbitrary likelihoods, limiting the accuracy of entropy models. Therefore, we use a discretized Gaussian Mixture Likelihoods to achieve a more flexible and accurate entropy model to enhance the performance. Besides, we utilize a simplified attention module with moderate complexity in our network architecture to achieve high coding efficiency.

Experimental results validate our proposed method achieves a state-of-the-art performance compared to existing learned compression methods and coding standards including HEVC, JPEG2000, JPEG. Besides, we achieve comparable performance with next-generation compression standard VVC for PSNR. The visual quality of our trained models using MS-SSIM outperforms existing methods.

\section*{Acknowledgement}

This work was supported in part by the Japan Society for the Promotion of Science (JSPS) Research Fellowship DC2 Grant Number 201914620, in part by JST, PRESTO Grant Number JPMJPR19M5, Japan.

\newpage
\onecolumn

\section{Appendix}

The section gives some ablation study, VVC settings and more results.

\subsection{Ablation Study on Network Architecture}

\subsubsection{Backbone}

In this paper, we use a similar structure as~\cite{IEEEexample:ourCLIC} in Fig.~\ref{fig:net}, whose original idea is from~\cite{IEEEexample:residual}. Four stacked $3\times3$ convolutions with residual connection can achieve larger receptive field with fewer parameters than one convolution with $5\times5$ kernel, which was used in the works~\cite{IEEEexample:Balle2, IEEEexample:David, IEEEexample:Lee}. On the other hand, subpixel convolution is used to replace commonly used transposed convolution as upsampling units to keep more details at the decoder side.

\begin{figure}[h]
\centering
\subfigure[Asymptotic performance]{
\label{Fig.netablation.1}
\includegraphics[height = 4.8cm]{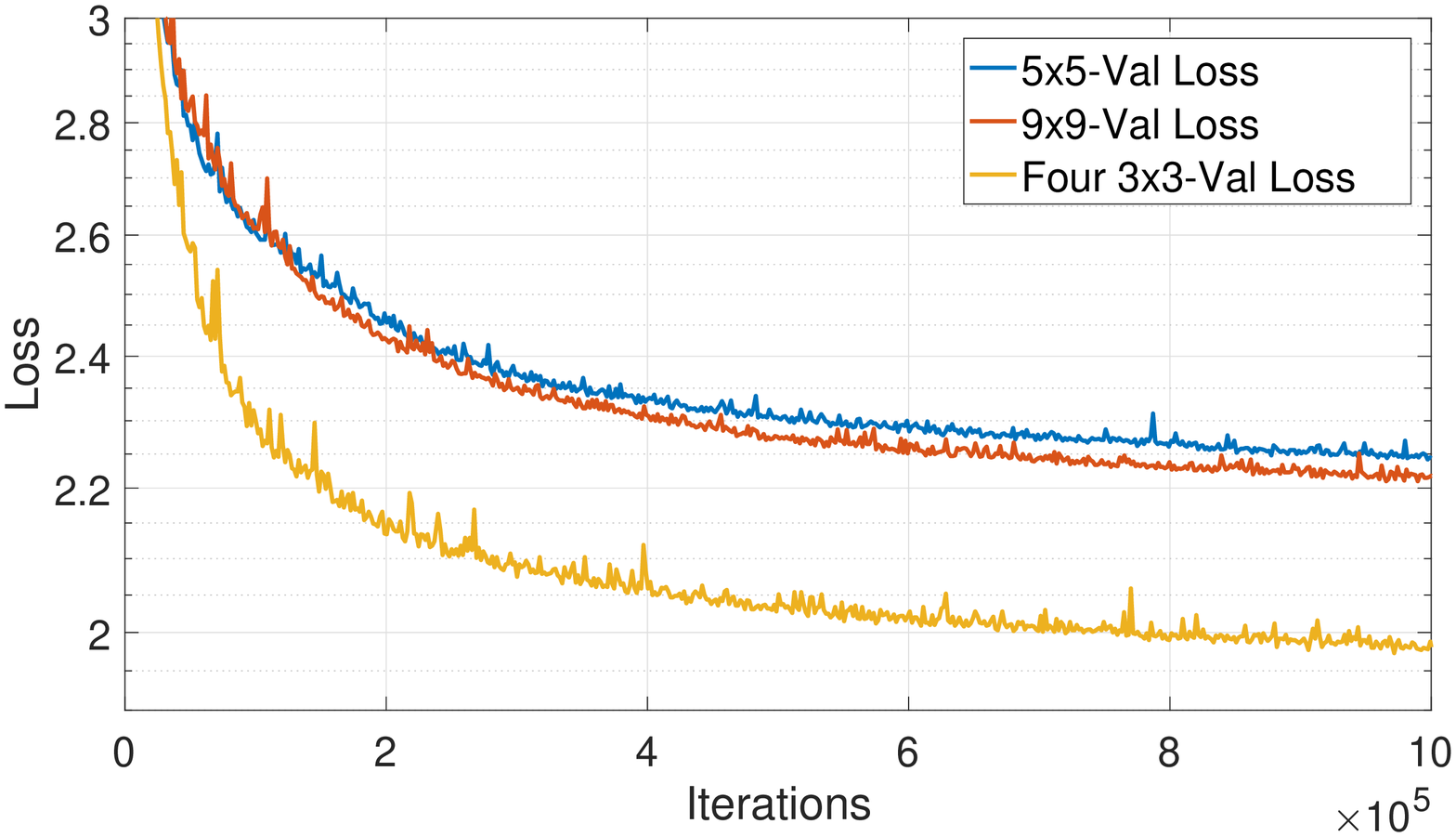}}
\subfigure[Rate-distortion performance at $1\times10^{6}$ iterations]{
\label{Fig.netablation.2}
\includegraphics[height = 4.8cm]{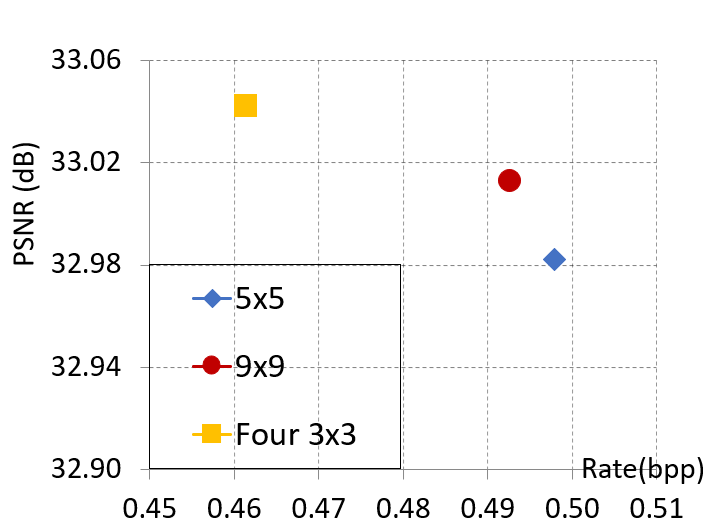}}
\caption{The ablation study on network architecture with $N=128$, optimized by MSE with $\lambda=0.015$.}
\label{fig:netablation}
\end{figure}

\begin{figure}[tb]
\centering
\subfigure[$5\times5$ kernel for each downsampling operation as~\cite{IEEEexample:Balle2, IEEEexample:David, IEEEexample:Lee}]{
\label{Fig.net.1}
\includegraphics[width=175mm]{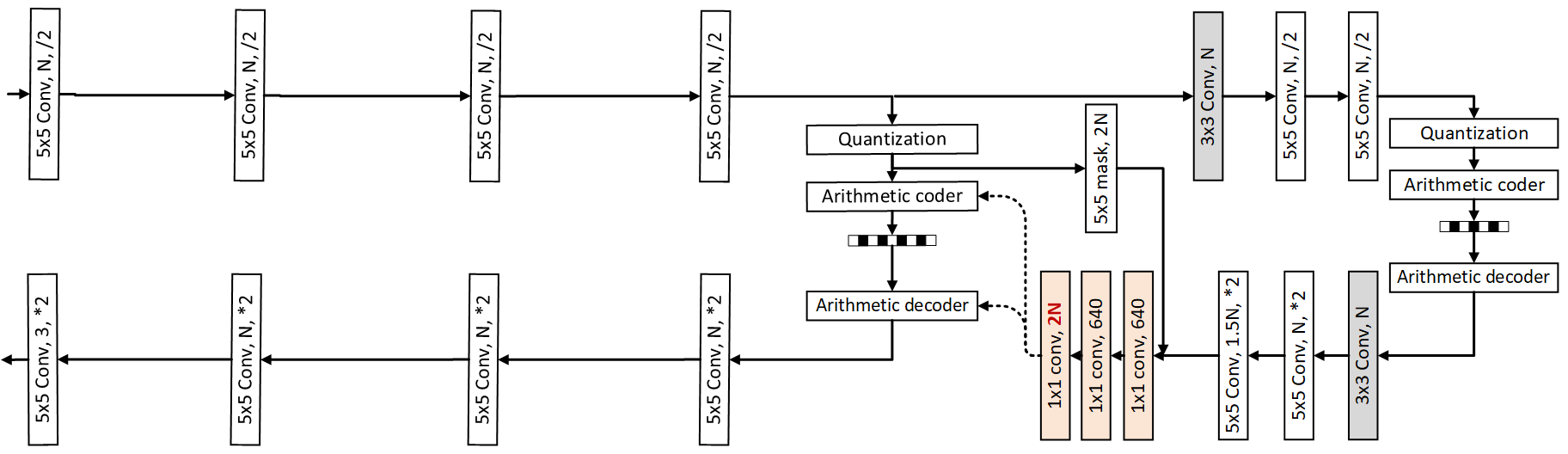}}
\subfigure[$9\times9$ kernel for each downsampling operation in the main autoencoder]{
\label{Fig.net.2}
\includegraphics[width=175mm]{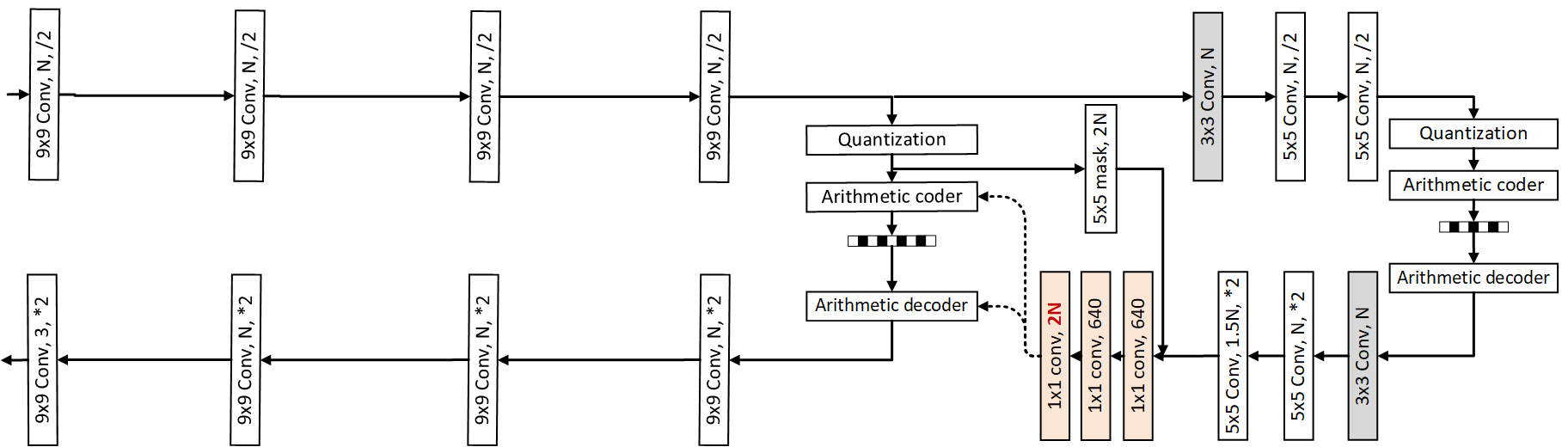}}
\subfigure[Four $3\times3$ kernels for each downsampling operation in the main autoencoder]{
\label{Fig.net.3}
\includegraphics[width=175mm]{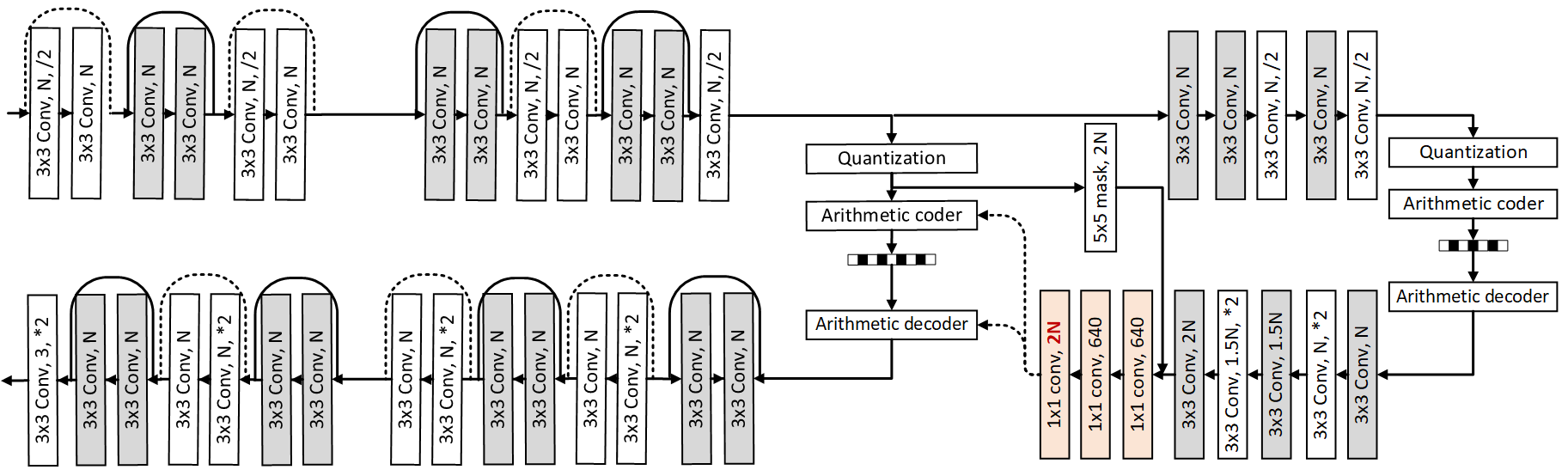}}
\caption{The ablation study on network architectures with single Gaussian entropy model ($K$=1).}
\label{fig:net}
\end{figure}

To illustrate the ablation study on network architecture, we have conducted experiments by comparing three cases: (a) $5\times5$ kernel for each downsampling operation as~\cite{IEEEexample:Balle2, IEEEexample:David, IEEEexample:Lee} as Fig.~\ref{Fig.net.1}; (b) $9\times9$ kernel for each downsampling operation in the main autoencoder as Fig.~\ref{Fig.net.2}. The kernel size in auxiliary autoencoder has fewer effect, so they are kept as $5\times5$;  (c) The network we used (denoted as our anchor) as Fig.~\ref{Fig.net.3}: four $3\times3$ kernels with residual connection for each downsampling operation and use subpixel convolution in synthesis transform. Except the network architecture, the other training settings are kept the same for the above three cases. The case (b) was tested, because it has the same architecture with (a), but has the same receptive field as (c). Gaussian mixture model is not incorporated, so single Gaussian model is used, requiring $2\times N$ channels for the output of entropy models. The number of filters $N$ is equal to $128$.

The loss curve and rate-distortion performance are shown in Fig.~\ref{fig:netablation}, respectively. It can be observed our network achieves better coding efficiency than the network architecture of~\cite{IEEEexample:Balle2, IEEEexample:David, IEEEexample:Lee}, by reducing the rate about 6\% (6\%=$\frac{0.49\text{bpp}-0.46\text{bpp}}{0.49\text{bpp}}$) with even slightly better quality. Therefore, we used the network architecture of Fig.~\ref{Fig.net.3} as the \emph{Anchor} in the Section 5.1 of our paper. One thing to note, the RD points in Figure 6(b) of original paper are slightly different from Fig.~\ref{Fig.netablation.2} only because all the RD points in Figure 6(b) are tested at about $4\times10^{5}$ iterations, and all the RD points in Fig.~\ref{Fig.netablation.2} are tested at about $1\times10^{6}$ iterations for fair comparison. The coding gain of our strategies always exist, and the asymptotic loss curves can illustrate this difference clearly.

\subsubsection{The Number of Mixtures $K$}

\begin{figure}[h]
\centering
\subfigure[Loss curves of single model and $K=2$]{
\label{Fig.mixture.1}
\includegraphics[height = 4.8cm]{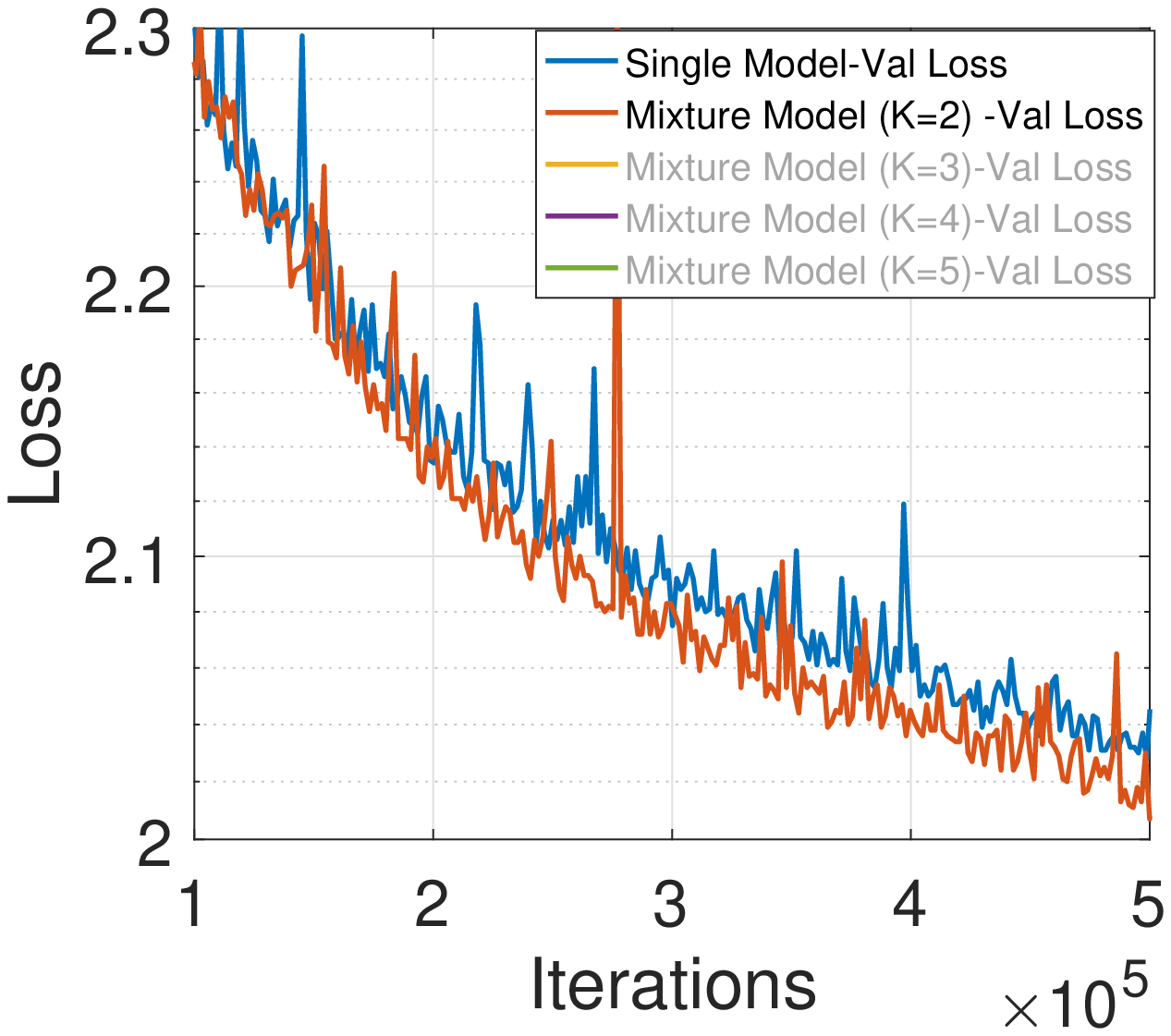}}
\subfigure[Loss curves of single model and $K=3$]{
\label{Fig.mixture.2}
\includegraphics[height = 4.8cm]{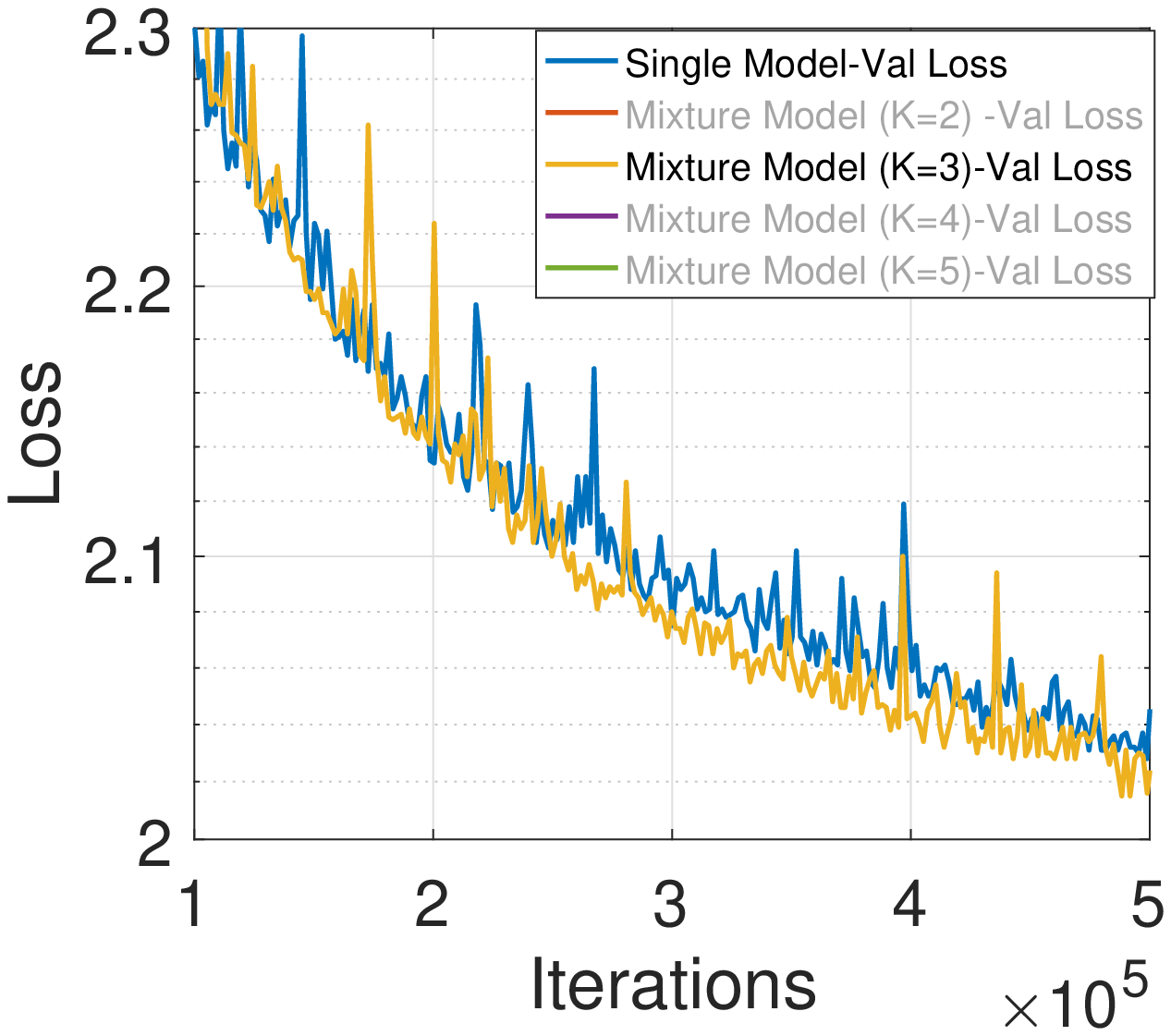}}
\subfigure[Loss curves of single model and $K=4$]{
\label{Fig.mixture.3}
\includegraphics[height = 4.8cm]{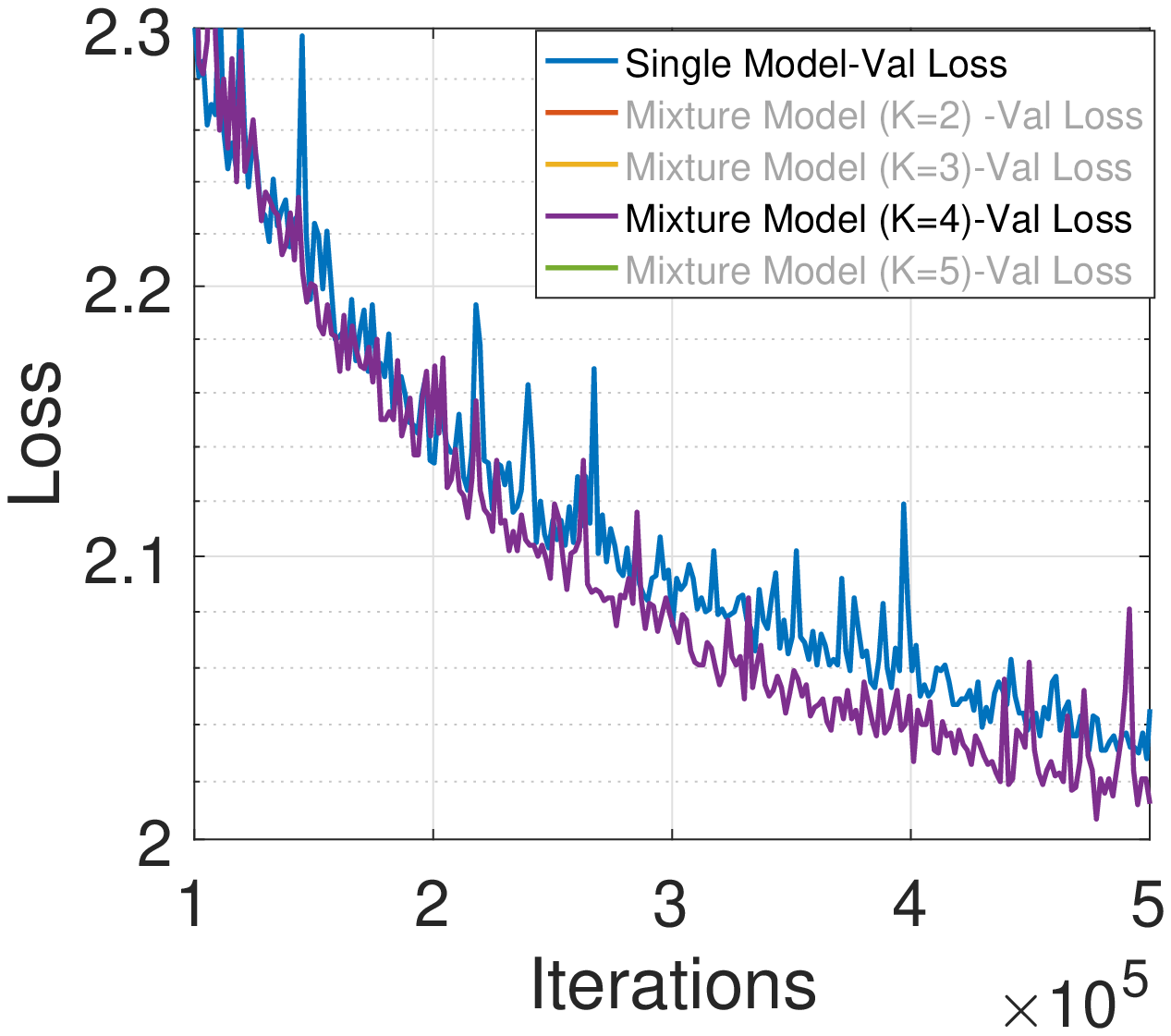}}
\subfigure[Loss curves of single model and $K=5$]{
\label{Fig.mixture.4}
\includegraphics[height = 4.8cm]{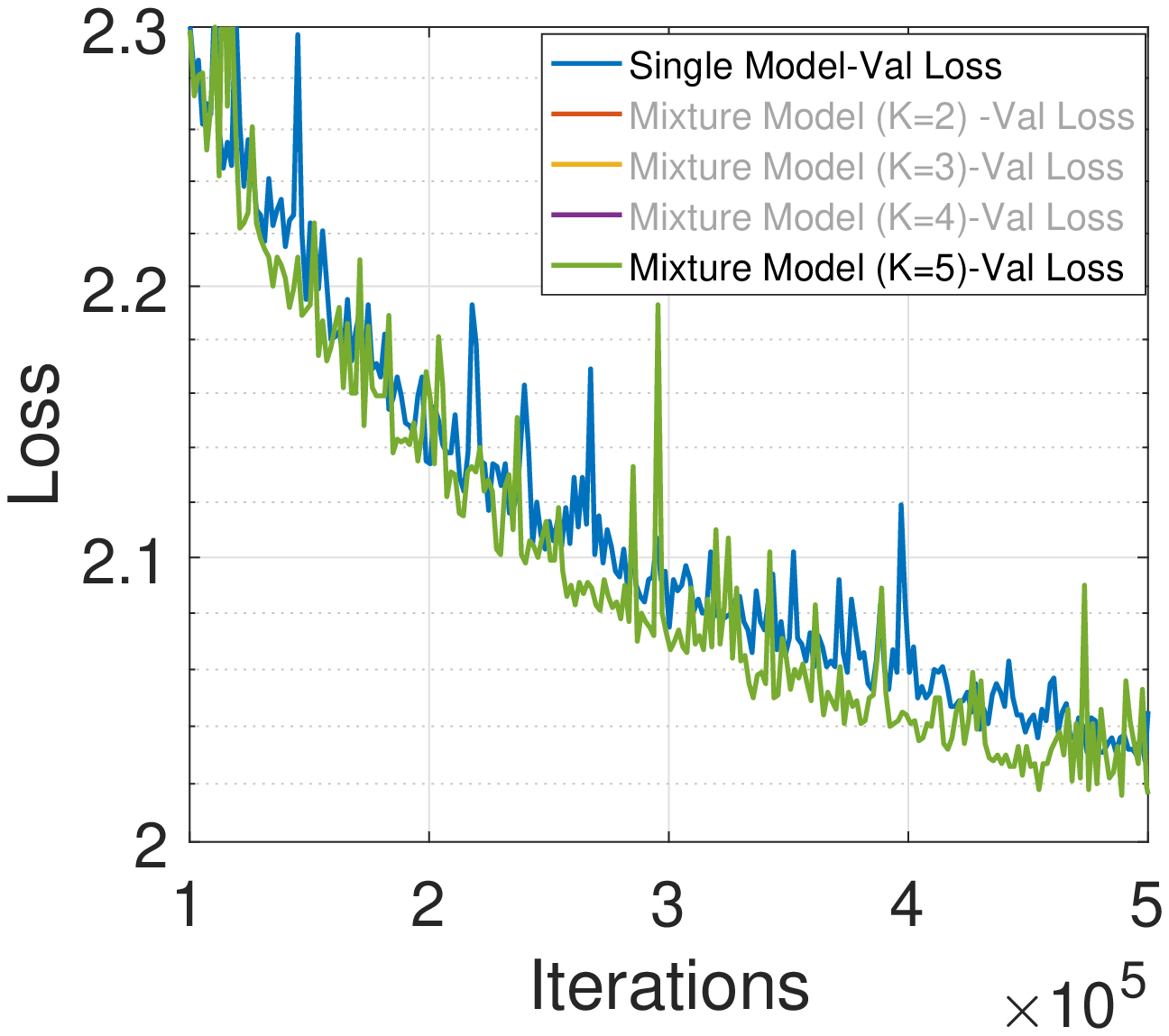}}
\subfigure[Rate-distortion performance]{
\label{Fig.mixture.5}
\includegraphics[height = 4.6cm]{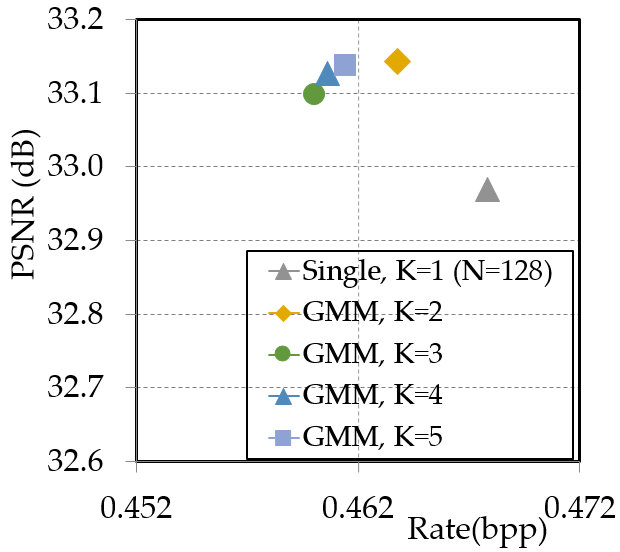}}
\caption{The ablation study on the number of mixtures $K$ with $N=128$, optimized by MSE with $\lambda=0.015$.}
\label{fig:mixture}
\end{figure}

In the paper, we used $K=3$ empirically. To validate the effect of the number of mixtures $K$, we tested the performance using K in the set of \{2, 3, 4, 5\} and results are discussed in Fig.~\ref{fig:mixture}. The loss values of mixture models are smaller than the loss of single Gaussian model, but the performance gain of different cases seems quite similar. Besides, we draw the RD points as shown in Fig.~\ref{Fig.mixture.5}. It can be observed that when K is equal to \{3, 4, 5\}, the performance almost saturates. We can not observe more coding gain by increasing the number of mixtures.

\begin{figure}[h]
\centering
\subfigure[Single Gaussian Model]{
\label{Fig.dist.1}
\includegraphics[width=160mm]{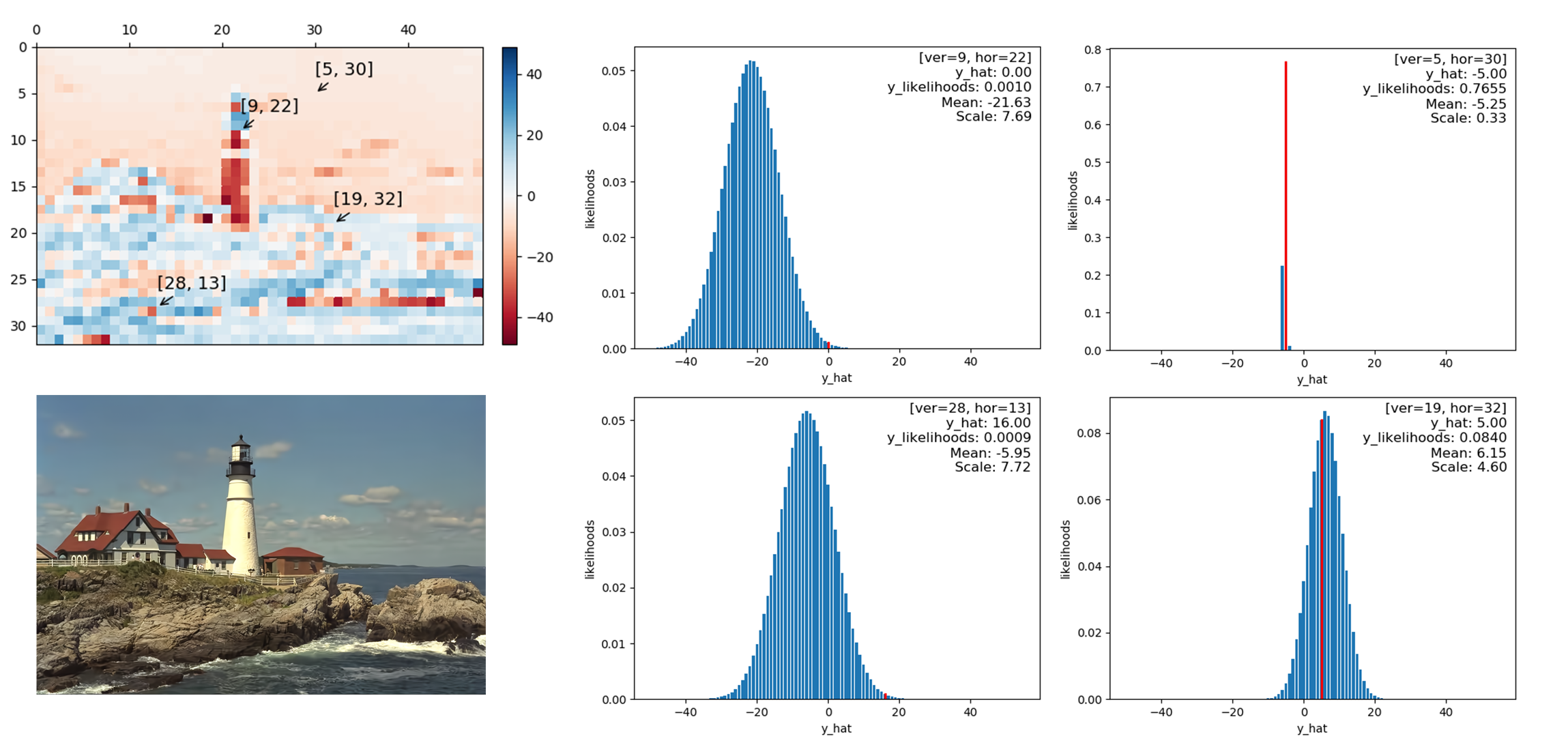}}
\subfigure[Gaussian Mixture Model]{
\label{Fig.dist.2}
\includegraphics[width=160mm]{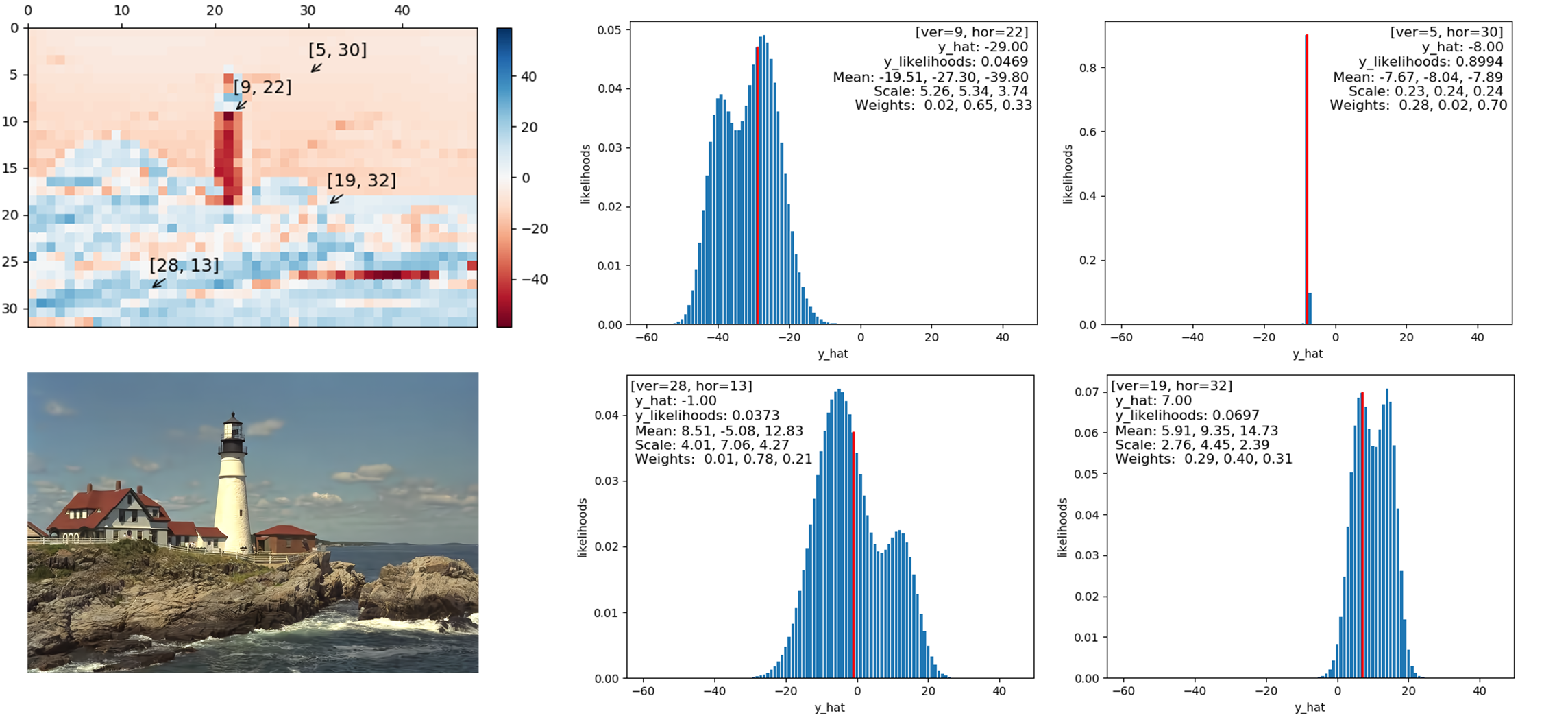}}
\caption{Visualization of estimated distributions for latent codes.}
\label{fig:dist}
\end{figure}

Moreover, to show the difference of single Gaussian distribution and mixture Gaussian distributions, we visualize the estimated likelihoods for some locations in the image \emph{Kodim21} from Kodak dataset in Fig.~\ref{fig:dist}. The size of \emph{Kodim21} is $512\times 768$, so the size of compressed codes $\boldsymbol{\hat{y}}$ is $32\times48$ after four downsampling operations in analysis transform. We select four representative locations (denoted as [Vertical axis, Horizontal axis]), that is, $[5, 30]$ in the sky, $[9, 22]$ at the white tower, $[28, 13]$ around the rock, $[19,32]$ between the boundaries. Single Gaussian model can only achieve symmetric and fixed shape in terms of discrete distribution, as shown in Fig.~\ref{Fig.dist.1}, while our proposed Gaussian mixture models achieves more flexible and arbitrary shapes in terms of discrete distribution, as shown in Fig.~\ref{Fig.dist.2}. It is noted that in many cases of simple regions, our model can degrade to single model distribution when three estimated mean values are the same, such as the location $[5, 30]$.

\subsection{Test Settings on Codecs}

\subsubsection{Versatile Video Coding (VVC) }

In order to test the performance of VVC, we used the VVC official test model VTM 5.2~\footnote{\url{https://vcgit.hhi.fraunhofer.de/jvet/VVCSoftware_VTM/tree/VTM-5.2}, accessed on July 17, 2019}. However, the dataset we used are RGB format, instead of YUV format, which are widely used in traditional compression standards. According to the document of VVC, given a RGB image, we convert RGB888 to YUV444 or YUV420 format using the definition from~\cite{IEEEexample:Color}. Then, the command line for compressing the given YUV files \emph{ImageYUV444.yuv} is

\vspace{3mm}
\;\; \emph{EncoderApp -c encoder\_intra\_vtm.cfg -i ImageYUV444.yuv -b ImageBinary.bin -o ImageRecon.yuv -f 1 -fr 2 -wdt ImageWidth -hgt ImageHeight -q QP -{}-OutputBitDepth=8 -{}-OutputBitDepth=8 -{}-OutputBitDepthC=8 -{}-InputChromaFormat=444} 
\vspace{3mm}

where \emph{encoder\_intra\_vtm.cfg} is the official intra configuration files by default. \emph{-f} denotes the number of frames to encode. VVC requires the frame rate must be larger than $1$, so we set \emph{-fr} as $2$, and we only have $1$ frame, so it did not affect the performance. \emph{-wdt} and \emph{-hgt} specify the image size. \emph{-q} specify the quantization parameters, and we use the QP in the set of $\{22, 27, 32, 37, 42, 47\}$. And all the YUV components have 8 bits. The color format is 444. On the other hand, YUV420 is more common than YUV444 in compression standards for a long history, so we also compress the image \emph{ImageYUV420.yuv} using the command line as

\vspace{3mm}
\;\; \emph{EncoderApp -c encoder\_intra\_vtm.cfg -i ImageYUV420.yuv -b ImageBinary.bin -o ImageRecon.yuv -f 1 -fr 2 -wdt ImageWidth -hgt ImageHeight -q QP -{}-OutputBitDepth=8 -{}-OutputBitDepth=8 -{}-OutputBitDepthC=8 -{}-InputChromaFormat=420} 
\vspace{3mm}

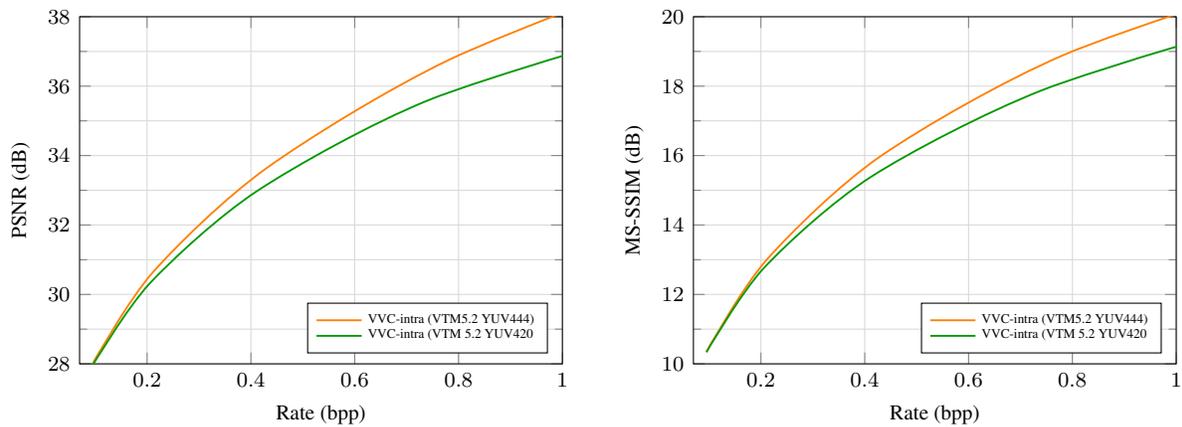
\begin{figure}[h]
\centering
\subfigure{
\pgfplotsset{
tick label style={font=\footnotesize},
label style={font=\footnotesize},
legend style={font=\footnotesize}
}
\pgfplotsset{every axis plot/.append style={line width=0.7pt}}
\tikzset{every mark/.append style={scale=0.5}}
\begin{tikzpicture}
\begin{axis}[
  width=8cm, height=6.2cm,
  xlabel = {Rate (bpp)},
  ylabel = {PSNR (dB)},
  xmin = 0.07,
  xmax = 1.0,
  ymin = 28,
  ymax = 38,
  minor y tick num = 1,
  grid = both,
  grid style = {gray!30},
  legend entries = {VVC-intra (VTM5.2 YUV444), VVC-intra (VTM 5.2 YUV420},
  legend style={font=\fontsize{5}{5}\selectfont, row sep=-3pt},
  legend cell align=left,
  legend pos = {south east},
  ]
  \addplot[smooth, orange] table {data_Kodak_VTM444_2.dat};
  \addplot[smooth, green!60!black] table {data_Kodak_VTM420_2.dat};

\end{axis}
\end{tikzpicture}
}
\subfigure{
\pgfplotsset{
tick label style={font=\footnotesize},
label style={font=\footnotesize},
legend style={font=\footnotesize}
}
\pgfplotsset{every axis plot/.append style={line width=0.7pt}}
\tikzset{every mark/.append style={scale=0.5}}
\begin{tikzpicture}
\begin{axis}[
  width=8cm, height=6.2cm,
  xlabel = {Rate (bpp)},
  ylabel = {MS-SSIM (dB)},
  xmin = 0.07,
  xmax = 1.0,
  ymin = 10,
  ymax = 20,
  minor y tick num = 1,
  grid = both,
  grid style = {gray!30},
  legend entries = {VVC-intra (VTM5.2 YUV444), VVC-intra (VTM 5.2 YUV420},
  legend style={font=\fontsize{5}{5}\selectfont, row sep=-3pt},
  legend cell align=left,
  legend pos = {south east},
  ]
  \addplot[smooth, orange] table {data_Kodak_VTM444.dat};
  \addplot[smooth, green!60!black] table {data_Kodak_VTM420.dat};
\end{axis}
\end{tikzpicture}
}
\caption{Performance Comparison of VVC codecs on Kodak dataset.}
\label{fig:vvc}
\end{figure}

The results are shown in Fig.~\ref{fig:vvc}. We can observe that the performance of YUV420 are worse than that of YUV444 due to some sampling loss of chroma components, especially decreasing the quality at high rate. Therefore, we used YUV444 for comparison in our paper.

\subsubsection{Boundary Handling}

For both traditional coding standards and deep learning based compression algorithms, boundary handling needs to be considered when the input images have arbitrary sizes. The CLIC validation images have arbitrary heights and widths. Therefore, for VVC the input size must be a multiple of the minimum CU size according to the specification of VVC encoder. Our solution is to pad the height and width of images to a multiple of $8$ using reflect padding. We also encode the height and width of input images into the bitstream and crop the required part to get the original size after decoding.

For our learned codec, the height and width of input images must be at least a multiple of $64$, because the minimum size in the networks are $[\frac{H}{64}, \frac{W}{64}]$ when the input size is $[H, W]$. Similarly, we pad the image height and width to a multiple of $64$ using reflect padding before feeding the data into the our learned neural network models, and encode the height and width into the bitstream. After the decoding, we crop the valid part to reconstruct the images with original size.

\end{document}